\documentclass[twocolumn]{aastex63}

\usepackage{amsmath}
\usepackage{accents}

\begin{document}

\title{\large{Global Spiral Density Wave Modes in Protoplanetary Disks: Morphology of Spiral Arms}}

\author[0000-0003-1277-6589]{\textcolor{black}{Enze Chen}}
\affiliation{Kavli Institute for Astronomy and Astrophysics, Peking University, Beijing 100871, China}
\affiliation{Department of Astronomy, School of Physics, Peking University, Beijing 100871, China} 

\author[0000-0002-3462-4175]{\textcolor{black}{Si-Yue Yu}}
\affiliation{Kavli Institute for Astronomy and Astrophysics, Peking University, Beijing 100871, China}
\affiliation{Department of Astronomy, School of Physics, Peking University, Beijing 100871, China}

\author[0000-0001-6947-5846]{\textcolor{black}{Luis C. Ho}}
\affiliation{Kavli Institute for Astronomy and Astrophysics, Peking University, Beijing 100871, China}
\affiliation{Department of Astronomy, School of Physics, Peking University, Beijing 100871, China}

\correspondingauthor{Si-Yue Yu}
\email{astroyue@outlook.com}

\begin{abstract}

We analyze two-armed global spiral density wave modes generated by gravitational instability in razor-thin, non-viscous, self-gravitating protoplanetary disks to understand the dependence of spiral arm morphology (pitch angle $\alpha$ and amplitude) on various disk conditions.  The morphologies of the resulting spiral density wave modes closely resemble observations. Their pitch angles and pattern speeds are insensitive to the boundary conditions adopted. Gaussian disks exhibit more tightly wound spirals (smaller pitch angle) than power law disks under the same conditions. We find that at a fixed disk-to-star mass ratio ($M_d/M_*$), pitch angle increases with average Toomre's stability parameter ($\overline Q$) or average disk aspect ratio ($\overline h$). For a given $\overline Q$, density wave modes with higher $M_d/M_*$ have larger pitch angles, while the behavior reverses for a given $\overline h$.  The interdependence between pitch angle and disk properties can be roughly approximated by $\alpha\propto c_s^2/M_d$, where $c_s$ is the sound speed.  
Our gravitational instability-excited spiral density waves can be distinguished from planet-launched spirals: (1) massive cool disks have spiral pitch angle falling with radius, while low-mass hot disks have spiral pitch angle rising with radius; (2) the profile of spiral amplitude presents several dips and bumps. We propose that gravitational instability-excited density waves can serve as an alternative scenario to explain the observed spiral arms in self-gravitating protoplanetary disks.

\end{abstract}

\keywords{protoplanetary disks}

\section{Introduction} \label{sec:introduction}

Substructures of protoplanetary disks, including cavities, gaps, rings, and spiral arms, have been revealed owing to detailed high-resolution observations \citep[e.g.,][]{ALMA2015,Dong2018a,Huang2018}. Spiral structure has been detected in protoplanetary disks in near-infrared scattered-light images that observe the disk surface \citep{Hashimoto2011, Muto2012, Garufi2013, Grady2013, Wagner2015, Avenhaus2017, Benisty2017, Canovas2018, Uyama2018}, in millimeter continuum images that probe the cold dust in the disk midplane  \citep{Perez2016, Huang2018, Kurtovic2018}, and in CO emission maps that detect the gas distribution \citep{Tang2012,Huang2020}. A significant fraction of observed spirals in protoplanetary disks are two-armed grand-design \citep[e.g.,][]{Garufi2013, Benisty2017, Huang2018}. Studies of the spiral arms and other substructures potentially help us understand the ongoing dynamic processes of protoplanetary disks \citep[e.g.,][]{Dipierro2015, Dong2015a, Dong2015b, Zhu2015}.

Gravitational perturbation from embedded planets or nearby stars drives spiral arms in protoplanetary disks. An embedded planet is expected to launch a primary one-armed spiral wake \citep{Goodman2001, Rafikov2002}. A secondary or even a tertiary spiral arm interior to the planet orbit arises \citep{Zhu2015, Miranda2019}, owing to constructive interference of different sets of density wave modes, with the arm number being dependent on the planet mass and disk temperature profile \citep{Bae2018}. Simulations of planet-disk interaction have produced planet-excited spiral wakes with appearance closely similar to observations \citep{Dong2015b, Dong2016b, Fung2015, Zhu2015}. Meanwhile, two-armed spirals also can be launched by a nearby star, such as those seen in the HD~100453 system \citep{Dong2016b, Wagner2018}. 

Another physical mechanism that can form spiral arms is gravitational instability \citep[GI; e.g.,][]{Rice2003, LodatoRice2005}.  GI occurs when the self-gravity that destabilizes the disk dominates over the restoring forces of thermal pressure and differential rotation, leading to redistribution of the disk material \citep[e.g.,][]{Toomre1964, Kratter2016}. In hydrodynamical simulations, protoplanetary disks with a cooling time scale longer than the dynamical time scale settle into a quasi-stable state, as cooling balances heating from viscous dissipation \citep{Rice2003,Cossins2009, Hall2019}. In contrast, cooling with shorter time scale would result in fragmentation, which grows into one or more gravitationally bounded objects \citep{Rice2003, Kratter2016}. Sufficiently massive disks never really settle down, possessing fewer and more short-lived arms whose amplitude changes strongly with time \citep{LodatoRice2005}. Hydrodynamical simulations based on GI can produce one, two, or more spiral arms \citep{Laughlin1994, Dipierro2014, Dong2015a, Hall2019}, in good agreement with observations of massive protoplanetary disks \citep{Meru2017, Tomida2017, Hall2018}.

The density wave theory \citep[][and references therein]{LinShu1964} predicts global-scale, GI-excited spiral density wave modes for self-gravitating unstable disks \citep{Bertin1989}. Density waves propagate through the disk and perturb the density distribution to form spiral structure \citep{LinShu1964, Binney2008, Shu2016}. Considering linear perturbations, \citet{Adams1989} derived one-armed density wave modes in protoplanetary disks, and \citet{Noh1991}, \citet{LaughlinRozyczka1996}, and \citet{Lee2019} calculated modes with two and more arms. The behavior of density wave modes is in good agreement with numerical simulations \citep{LaughlinRozyczka1996, Laughlin1998}. Theoretical studies have been carried out to analyze the mode properties, including non-linear evolution \citep{Papaloizou1991}, energy and angular momentum transport \citep{Lin2015}, mode-mode interaction \citep{LaughlinKorchagin1996}, and mode saturation \citep{Laughlin1997}.

The degree of spiral arm winding is described by the pitch angle ($\alpha$), which is defined as the  the angle between the tangent of the spiral and the azimuthal direction \citep{Binney2008}. Large/small pitch angle corresponds to loosely/tightly wound spiral arms. It has been shown that the pitch angle of planet-excited spiral density waves becomes larger (more loosely wound) for disks with higher sound speed, disks associated with a more massive perturber \citep{Rafikov2002, Zhu2015}, and disks with longer cooling time scale \citep{ZhangZhu2020}. In addition, the arm pitch angle increases with radius for arms interior to the perturber orbit, with the trend reversing  for arms exterior to this orbit \citep{Rafikov2002}.  The arms exterior to the perturber orbit are very tightly wound \citep{Dong2015b}. 

Within this backdrop, the behavior of pitch angles of GI-excited density wave modes remains somewhat unclear. Understanding such a behavior would promote our understanding of spiral arms in protoplanetary disks and may provide us a tool to distinguish the different formation mechanisms of observed spirals.  In this work, we calculate two-armed spiral density wave modes in protoplanetary disks with various disk conditions, aiming to investigate the interdependence between their pitch angles and protoplanetary disk properties. We describe our analytical model and numerical solution method in Section~\ref{sec:method}.  Spiral density wave modes are calculated in Section~\ref{sec:results}.  The implications of our results are discussed in Section~\ref{sec:discussion}. Section~\ref{sec:summary} summarizes our findings. 

\section{Method} \label{sec:method}

We numerically solve the equations of fluid dynamics to compute two-armed global spiral density wave modes in protoplanetary disks in the framework of linear density wave theory. The calculations employ a two-dimensional, razor-thin, non-viscous fluid disk, taking into consideration the kinematic effects of the central star's gravity, disk self-gravity, and thermal pressure. 

\subsection{Linear Dynamics Equations}
\citet{Adams1989} were the first to lay out the linear perturbation equations and derive one-armed density wave modes within the context of protostellar disks. We build our model based on the formalism of \citet{Adams1989}, and conform to their notation system. We use cylindrical coordinates $(r,\phi,z)$ and restrict the disk to the $z=0$ plane. The disk dynamics are characterized by the  continuity equation (Equation~\ref{eq:continuity}), the radial and azimuthal momentum equations (Equation~\ref{eq:momentum}), Poisson's equation (Equation~\ref{eq:Possion}), and the equation of state (Equation~\ref{eq:EOS}):  

\begin{equation}    \label{eq:continuity}     
\frac{\partial\sigma}{\partial t}+\frac{1}{r}\frac{\partial(r\sigma u)}{\partial r}+\frac{1}{r}\frac{\partial(\sigma v)}{\partial\phi}=0, 
\end{equation}

\begin{subequations}  \label{eq:momentum}
\begin{equation} 
\frac{\partial u}{\partial t}+u\frac{\partial u}{\partial r}+\frac{v}{r}\frac{\partial u}{\partial \phi}-\frac{v^2}{r}=-\frac{GM_*}{r^2} -\frac{\partial(\Psi+\zeta)}{\partial r}, 
\end{equation}

\begin{equation}  
\frac{\partial v}{\partial t}+u\frac{\partial v}{\partial r}+\frac{v}{r}\frac{\partial v}{\partial \phi}+\frac{vu}{r}=-\frac{1}{r}\frac{\partial (\Psi+\zeta) }{\partial \phi}, 
\end{equation}
\end{subequations}

\begin{equation}  \label{eq:Possion}      
\nabla^2\Psi=4\pi G\sigma \delta(z), 
\end{equation}

\begin{equation} \label{eq:EOS} 
\mathrm{d}\zeta=c_s^2\frac{\mathrm{d}\sigma}{\sigma}, 
\end{equation}

\noindent
where $u(r, \phi, t)$ denotes the radial velocity, $v(r, \phi, t)$ the azimuthal velocity, $M_*$ the central star mass, $c_s$ the sound speed, following $c_s^2=\mathrm{d}p/ \mathrm{d}\sigma\propto T$, $\sigma(r, \phi, t)$ the  surface density, $\Psi(r, \phi, t)$ the disk self-gravity potential, and $\zeta(r, \phi, t)$ the enthalpy. The enthalpy $\zeta$ is used instead of the thermal pressure $p$ for convenience of the calculation. We assume that the disk is in an adiabatic state, a widely adopted assumption in protoplanetary disk modeling \citep[see][]{LaughlinRozyczka1996,Miranda2019}.

For linear analysis, we assume that the spiral perturbation is small compared with the disk background, and hence can be Fourier-analyzed in time $t$ and azimuthal angle $\phi$. All the variables ($u, v, \sigma, \Psi, \zeta$) can then be decomposed as 

\begin{equation}  \label{eq:Fourier summation decomposition}   
F(r,\phi,t)=F_0(r)+\sum_{m=1}^{\infty} F_1^m(r)e^{i(\omega_m t-m\phi)}, 
\end{equation} 

\noindent where the azimuthal wavenumber $m$ indicates the number of spiral arms. We focus on two-armed ($m=2$) spiral density wave modes, because of the prevalence of two-armed spirals among current observations: in near-infrared scattered-light observations, a significant fraction of observed spirals have two arms \citep[e.g.,][]{Garufi2013, Benisty2015, Benisty2017, Canovas2018, Uyama2018};  millimeter continuum imaging has revealed six protoplanetary disks containing spiral structures, five of which have symmetric two-armed spirals \citep{Andrews2018, Huang2018, Kurtovic2018}. Meanwhile, massive gravitationally unstable disks, as probed in this work, tend to have low-order modes in simulations \citep{Dong2015a, Hall2019}. Higher order density waves, however, are more likely to be absorbed at the Lindblad resonance radius (see Sections~\ref{subsec:existence} and \ref{subsec:caveats} for further discussion).

Focusing on the $m=2$ mode, Equation~(\ref{eq:Fourier summation decomposition}) is reduced to

\begin{equation}  \label{eq:linear decomposition}   
F(r,\phi,t)=F_0(r)+F_1(r)e^{i(\omega t-m\phi)}.
\end{equation}
\

\noindent The equilibrium states $F_0(r)$ of the physical properties of our protoplanetary disk models (disk mass, density profile, temperature profile, etc.) are discussed and determined in Section~\ref{subsec:disk properties}. We aim to numerically compute the spiral perturbation $F_1(r)$ and the wave frequency $\omega$. The real part of $F_1(r)e^{i(\omega t-m\phi)}$ is physically meaningful. The perturbation density $\sigma_1$ gives the density profile of the global spiral density wave modes. The wave frequency $\omega$ is defined as $\omega=m\Omega_p-i\gamma$. The pattern speed $\Omega_p$ specifies the angular velocity of the steadily rotating spiral pattern, and $\gamma$ gives the exponential growth rate of the spiral mode. Considering quasi-steady spirals with steady morphology for at least several dynamical time scales, the mode has a constant wave frequency, i.e., $\Omega_p$ and $\gamma$, that does not vary with radius and azimuthal angle in one disk \citep{Bertin1989}. $\Omega_p$ and $\gamma$ are free parameters and are determined by the eigenstate of the integro-differential equation (Equation \ref{eq:int-diff}).

After linearization, Equations (\ref{eq:continuity})--(\ref{eq:EOS}) become

\begin{equation} \label{eq:continuity linearized}   
i(\omega-m\Omega)\frac{\sigma_1}{\sigma_0}+\left(\frac{1}{r}+\frac{1}{\sigma_0}\frac{\mathrm{d}\sigma_0}{\mathrm{d}r}\right)u_1+\frac{\mathrm{d}u_1}{\mathrm{d}r}-\frac{im}{r}v_1=0,
\end{equation}

\begin{subequations}  \label{eq:momentum linearized}
\begin{equation} 
i(\omega-m\Omega)u_1-2\Omega v_1=-\frac{\mathrm{d}(\Psi_1+\zeta_1)}{\mathrm{d}r},
\end{equation}

\begin{equation}  
i(\omega-m\Omega)v_1+\frac{\kappa^2}{2\Omega}u_1=\frac{im}{r}(\Psi_1+\zeta_1),
\end{equation}
\end{subequations}

\begin{equation}  \label{eq:potential integration} 
\Psi_1(r)=-\int_{R_{\rm inn}}^{R_{\rm out}} \mathrm{d}\rho \int_0^{2\pi} \frac{G\cos(m\phi)\sigma_1(\rho)\rho}{\sqrt{\rho^2+r^2-2\rho r\cos\phi}}\mathrm{d}\phi,
\end{equation}

\begin{equation} \label{eq:EOS linearized} 
\zeta_1=c_s^2\frac{\sigma_1}{\sigma_0}, 
\end{equation}

\noindent
where subscript 0 denotes the equilibrium state and subscript 1 denotes spiral perturbation quantities. The angular velocity of the equilibrium state is

\begin{equation}  \label{eq:omega}  
\Omega=\sqrt{\frac{1}{r}\left(\frac{GM_*}{r^2}+\frac{\mathrm{d}\Psi_0}{\mathrm{d}r}+\frac{\mathrm{d}\zeta_0}{\mathrm{d}r} \right)},
\end{equation}

\noindent
with epicyclic frequency

\begin{equation}  \label{eq:kappa} 
\kappa^2=\frac{1}{r^3}\frac{\mathrm{d}}{\mathrm{d}r}[(r^2\Omega )^2].
\end{equation}
\

Combining Equations (\ref{eq:continuity linearized})--(\ref{eq:EOS linearized}), we eliminate other variables and get a second-order integro-differential equation of the perturbation density $\sigma_1(r)$ and wave frequency $\omega$: 

\begin{equation} \label{eq:int-diff} 
\left[\frac{\mathrm{d}^2}{\mathrm{d}r^2}+A(\omega,r)\frac{\mathrm{d}}{\mathrm{d}r}+B(\omega,r) \right](\zeta_1+\Psi_1)+C(\omega,r)\zeta_1=0, 
\end{equation}

\noindent
where the coefficients $A$, $B$, and $C$ are given by

\begin{equation}  
A=\frac{\mathrm{d}}{\mathrm{d}r}\ln \left[ \frac{\sigma_0r}{\kappa^2(1-\nu^2)} \right],
\end{equation}

\begin{equation}  
B=-\frac{m^2}{r^2}-\frac{4m\Omega}{\kappa (1-\nu^2)r}\frac{\mathrm{d}\nu}{\mathrm{d}r}+\frac{2m\Omega}{\kappa\nu r}\frac{\mathrm{d}}{\mathrm{d}r}\ln\left(\frac{\kappa^2}{\Omega \sigma_0} \right),
\end{equation}

\begin{equation} 
C=-\frac{\kappa^2(1-\nu^2)}{c_s^2}.
\end{equation}

\noindent
In these equations, the Doppler-shifted frequency is defined as

\begin{equation}  
\nu=\frac{\omega-m\Omega}{\kappa}.
\end{equation}
\

Together with the boundary conditions defined in Section~\ref{subsec:BC}, Equation~(\ref{eq:int-diff}) becomes a well-defined integro-differential equation. We then numerically solve Equation~(\ref{eq:int-diff}) without further approximation and obtain $\sigma_1(r)$ and $\omega$ (Section~\ref{subsec:numerical method}).

\subsection{Boundary Condition} \label{subsec:BC}

Following the strategy of \cite{Noh1991}, we adopt two different inner boundary conditions: reflecting (R) and transmitting (T). For a reflecting boundary condition, the radial velocity is set to vanish ($u_1=0$), resulting in

\begin{equation} \label{eq:reflect}
 \left(\frac{\mathrm{d}}{\mathrm{d}r}-\frac{2m\Omega}{\omega-m\Omega}\frac{1}{r}  \right)(\Psi_1 +\zeta_1)=0.
\end{equation}

\noindent
For a transmitting boundary condition, we assume that the trailing waves propagate through disk boundaries, which requires 

\begin{equation}
	\mathrm{d}\sigma_1/\mathrm{d}r + i|k|\sigma_1=0, 
\end{equation}

\noindent
with the wave number $|k|$ obtained from the dispersion relation (see Equation~\ref{eq:dispersion relation})

\begin{equation}
|k|=\frac{\pi G \sigma_0}{c_s^2}+\frac{\pi G \sigma_0}{c_s^2} \{1-Q^2[1-(\omega-m\Omega)^2/\kappa^2]\}^{1/2}.
\end{equation}

We adopt three different outer boundary conditions: confining pressure boundary condition (C), as well as the reflecting and transmitting boundary conditions. Confining pressure is set where the disk is conjuncted with the outer environment, leading to a constant background pressure. Requiring that the Lagrangian pressure perturbation $\Delta p=0$, and hence $\Delta \sigma=0$, 

\begin{equation} \label{eq:confining}
\sigma_1=\frac{1}{\kappa^2-(\omega-m\Omega)^2}\frac{\mathrm{d}\sigma_0}{\mathrm{d}r} \left(\frac{\mathrm{d}}{\mathrm{d}r}-\frac{2m\Omega}{\omega-m\Omega}\frac{1}{r}  \right)(\Psi_1+\zeta_1).
\end{equation}

\noindent
As discussed in Section~\ref{subsec: boundary results}, the boundary conditions influence the wave behavior near the boundaries and the wave feedback process.

\subsection{Numerical Method}  \label{subsec:numerical method}

We solve the integro-differential Equation (\ref{eq:int-diff}) using the matrix-eigenvalue method described in \citet{Adams1989}. We briefly summarize the key steps and leave additional details to Appendix~\ref{app:numerical}. In this method, Equation~(\ref{eq:int-diff}) eventually turns into an eigenvalue problem, with $\omega$ and $\sigma_1$ being the eigenfrequency and eigenfunction. This eigenvalue method has been widely adopted in searching for global normal modes in planetary rings, galaxy disks, and protostellar disks \citep[e.g.,][]{Papaloizou1989, Noh1991, LaughlinRozyczka1996, Feng2014, Lee2019}. 

We set $N=500$ logarithmically spaced radial grid points to discretize Equation~(\ref{eq:int-diff}), aiming to solve the value of $\sigma_1(r)$ at each grid point and $\omega$ for each eigenmode. For convenience, our calculation uses dimensionless $S(r)=\sigma_1(r)/\sigma_0(r)$ instead of $\sigma_1(r)$.  $S(r)$ is then replaced by the $N$-dimensional vector \emph{\textbf{S}}, which contains $N$ values of $S(r)$ at each grid point. Likewise, all the functions and operators in Equation~(\ref{eq:int-diff}) are replaced by $N\times N$ matrices (Appendix~\ref{app:numerical}), turning it into the matrix equation

\begin{equation}      \label{eq:int-diff matrix}  
\mathcal{W}(\omega)\emph{\textbf{S}}=0.
\end{equation}

\noindent
The $N\times N$ matrix $\mathcal{W}(\omega)$ is a fifth-order function of $\omega$. We analytically regroup Equation~(\ref{eq:int-diff matrix}) into the $5N$-dimensional 

\begin{equation}      \label{eq:eigen}  
\mathcal{T}\emph{\textbf{S}}_\emph{\textbf{d}}=\omega \mathcal{Z} \emph{\textbf{S}}_\emph{\textbf{d}},
\end{equation}
 
\noindent
which is a standard eigenvalue problem, where $\emph{\textbf{S}}_\emph{\textbf{d}}$ is analytically equivalent to \emph{\textbf{S}}. We refer the reader to Appendix~\ref{app:numerical} for the detailed transformation process.  Solving the eigenfunction and eigenfrequency of Equation (\ref{eq:eigen}) gives the perturbation density \emph{\textbf{S}} and wave frequency $\omega$. Equation~(\ref{eq:eigen}) has $5N$ eigenmodes, only a few of which have non-zero growth rate. We choose the eigenmode with the highest growth rate as our final spiral density wave mode, as it would grow exponentially faster and dominate over the other modes \citep{Adams1989, Bertin1989}. We note that linear analysis is unable to give the amplitude of $\sigma_1$, as $\sigma_1$ can be multiplied by an arbitrary factor, as seen in Equation~(\ref{eq:int-diff}). Information on the amplitude is lost when $F_1\ll F_0$ is assumed. In our numerical calculation, we employ a unit system in which the outer disk radius $R_{d}$, the central star mass $M_*$, the gravitational constant $G$, and the Keplerian angular velocity at the disk outer boundary ($\Omega_{\rm K}=G^{1/2}M_*^{1/2} \cdot R_{d}^{-3/2}$) are set to unity. The angular velocity, pattern speed, and growth rate in our calculation have the units of $\Omega_{\rm K}$. 

The two-dimensional gravitational potential calculation has a singularity that needs special care.  To deal with the singularity, we soften the gravity by adding a term $\eta^2r^2$ with $\eta\ll1$ \citep{Laughlin1997, Tremaine2001} to the denominator of Equation~(\ref{eq:potential integration}) (for details, see Appendix~\ref{app:numerical}).

\section{Results}  \label{sec:results}

We numerically calculate two-armed spiral density wave modes in protoplanetary disks of various physical properties. We use pitch angle to characterize the degree of arm winding, and then study the interdependence between spiral pitch angles and disk properties.

\subsection{Disk Properties}  \label{subsec:disk properties}

To probe the spiral density wave modes generated by GI, we consider massive protoplanetary disks with disk-to-star mass ratio ranging from $M_d/M_* \approx 0.1$ to $0.5$. These disks are sufficiently massive that the disk self-gravity plays an important role. It has been shown, based on both theory and simulations, that GI in protoplanetary disks occurs when $M_d/M_*\gtrsim 0.1$ \citep{Adams1989, Noh1991, Dong2015a, Hall2018}. In contrast, observations give  typical disk masses of $M_d/M_*\approx 0.01$ \citep{Andrews2013, Dong2018b}. Disk masses derived from current observations, however, have large uncertainties and may be underestimated \citep{Kratter2016,Dong2018a}. In addition, massive disks may exist in an earlier evolutionary stage \citep{Kratter2016}. Section~\ref{subsec:existence} presents further discussion.

GI is quantified by Toomre's (1964) parameter $Q = \kappa c_s/\pi G \sigma_0$. A smaller $Q$ generally indicates a more unstable disk. The condition that $Q\lesssim 1$ is the threshold of axisymmetric instability, while the threshold value of $Q$ is greater than unity for non-axisymmetric spiral perturbations \citep{Kratter2016}. We utilize the mass-weighted average value of $\overline Q= \int Q\sigma_0 \mathrm{d}S/\int \sigma_0 \mathrm{d}S$, where $S$ denotes the surface area, to characterize the disk stability, and focus on marginally stable disks with $1 \lesssim \overline Q \lesssim 2$. In this case, the disks are stable to axisymmetric perturbation but unstable to non-axisymmetric spiral perturbation, as shown in our resulting density wave modes. 

\begin{figure}
\plotone{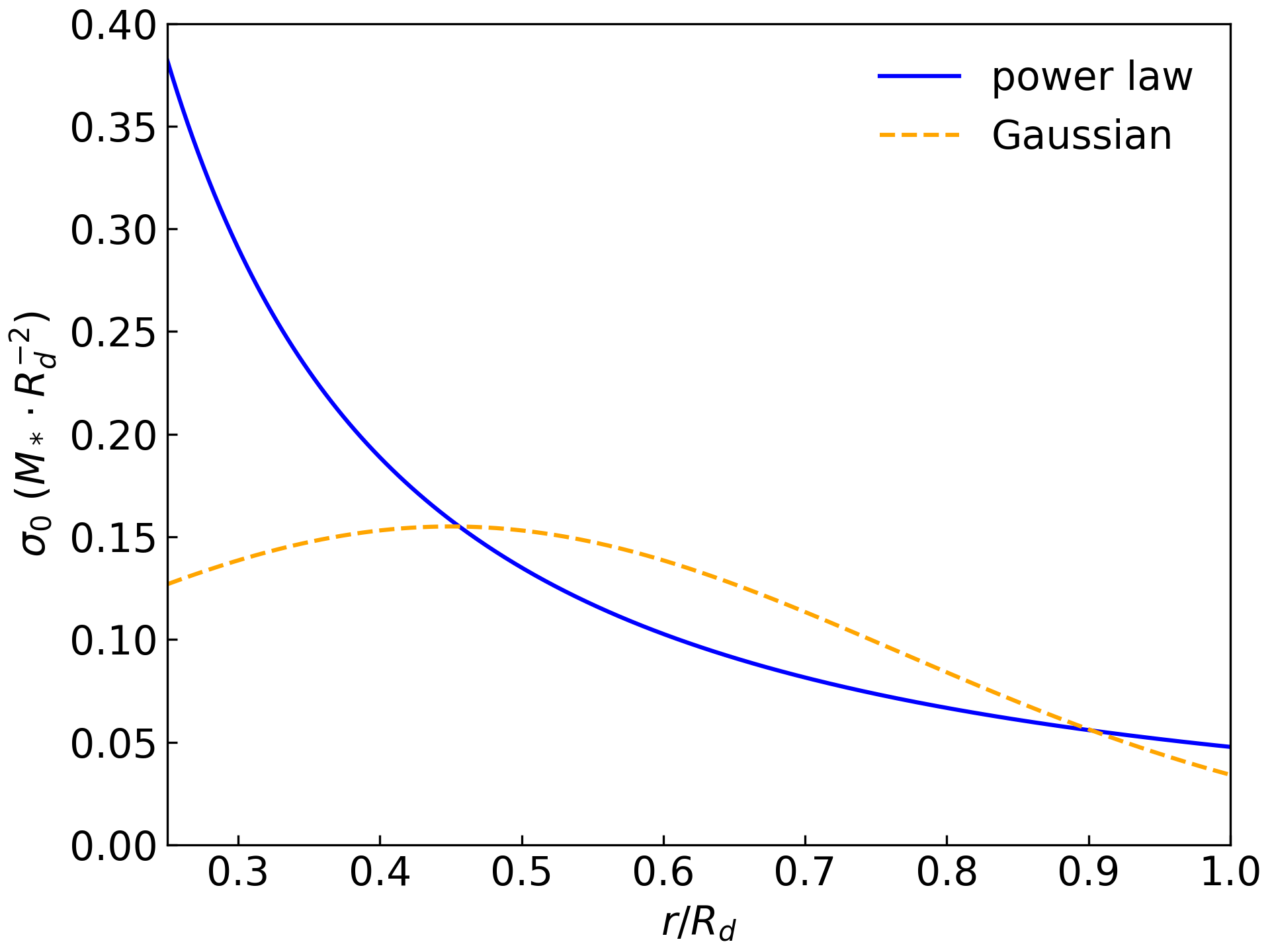}
\caption{Surface density profiles of the fiducial power law disk ($\sigma_0=\Sigma^{\rm p} r^{-1.5}$; blue solid line) and the fiducial Gaussian disk ($\sigma_0=\Sigma^{\rm G} e^{-(r-0.45)^2/0.2}$; orange dashed line) with $M_d/M_*=0.3$. \label{Fig:density contrast}}
\end{figure}

We make use of two kinds of surface density profiles: power law disk and Gaussian disk.  The power law density profile follows $\sigma_0(r)=\Sigma\, r^{-p}$, which is widely adopted in both observational \citep{Wilner2000, Pietu2006, Andrews2009, Grady2013} and theoretical \citep{Papaloizou2002, Dong2015a, Hall2018} studies. The value of $p$ ranges from 0 to 2. The Gaussian profile, $\sigma_0(r)=\Sigma\, e^{-(r-R_0)^2/w}$, characterizes more evolved disks with inner disk region beginning to be depleted \citep[e.g.,][]{Laughlin1997, Boehler2017, Pinilla2018}. We adopt $p=1.5$, $R_0=0.45$, and $w=0.2$ in our models. The normalization factor $\Sigma$ is determined by the disk mass. Figure~\ref{Fig:density contrast} illustrates the density profiles of the fiducial disk for each kind.

%===================================================
%===================================================

We utilize a power law temperature profile $T_0(r)=T_{k}\, r^{-q}$, which is widely used in fitting observations \citep{Andrews2011, Muto2012, Dullemond2020}, $q \approx 0.1-1$. For a disk in which cooling by self-radiation balances heating from irradiation of the central star, the temperature profile, depending on disk flaring, has a varying $q$, which typically equals $3/7$ \citep{Chiang1997, Kratter2016}. If cooling balances heating by the release of gravitational potential energy via accretion, the temperature profile has $q=3/4$ \citep{Kratter2016}. Our models adopt an intermediate value of $q=0.6$. The aspect ratio $h=c_s/(\Omega r)$ characterizes the disk thickness. As $\Omega\propto r^{-1.5}$, we have $h\propto r^{0.2}$ in our models, indicating a nearly constant flare angle. The mass-weighted average aspect ratio is defined as $\overline h= \int h\sigma_0 \mathrm{d}S/\int \sigma_0 \mathrm{d}S$. The normalized factor $T_{k}$ in the temperature profile is determined by $\overline Q$ or $\overline h$ through the sound speed $c_s$.

The radial extent of our disk models is set from 25\,AU to 100\,AU. The region near the central star is excluded, considering that central disk region can be depleted and involve non-linear processes \citep{Andrews2011}. The influence of boundary conditions on the resulting density wave modes was tested (discussed in Section~\ref{subsec: boundary results}). Spiral wave modes in disks with different boundaries generally show similar morphology and pitch angle. In our following calculations, we use two representative groups of boundary conditions: (1) reflecting inner boundary and transmitting outer boundary for power law disk, and (2) reflecting inner boundary and confining pressure outer boundary for Gaussian disk. Current observations provide little constraint on the physical properties of the disk boundaries. We choose these two groups in order to consider both reflecting and transmitting density waves in the disks.

\subsection{Fiducial Disks}

%---------------------------------------------------------------------
\begin{figure*}[h!]
\plottwo{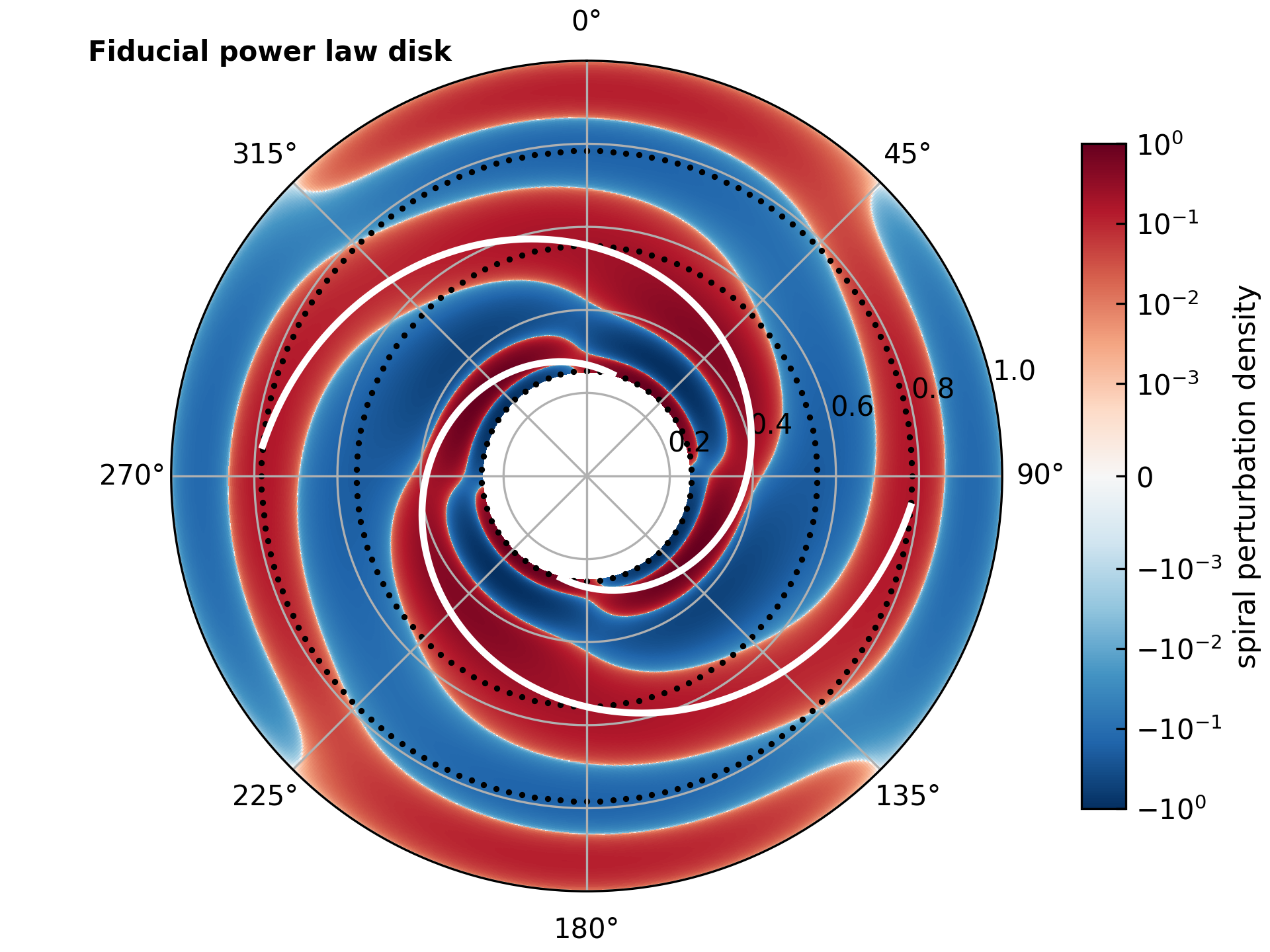}{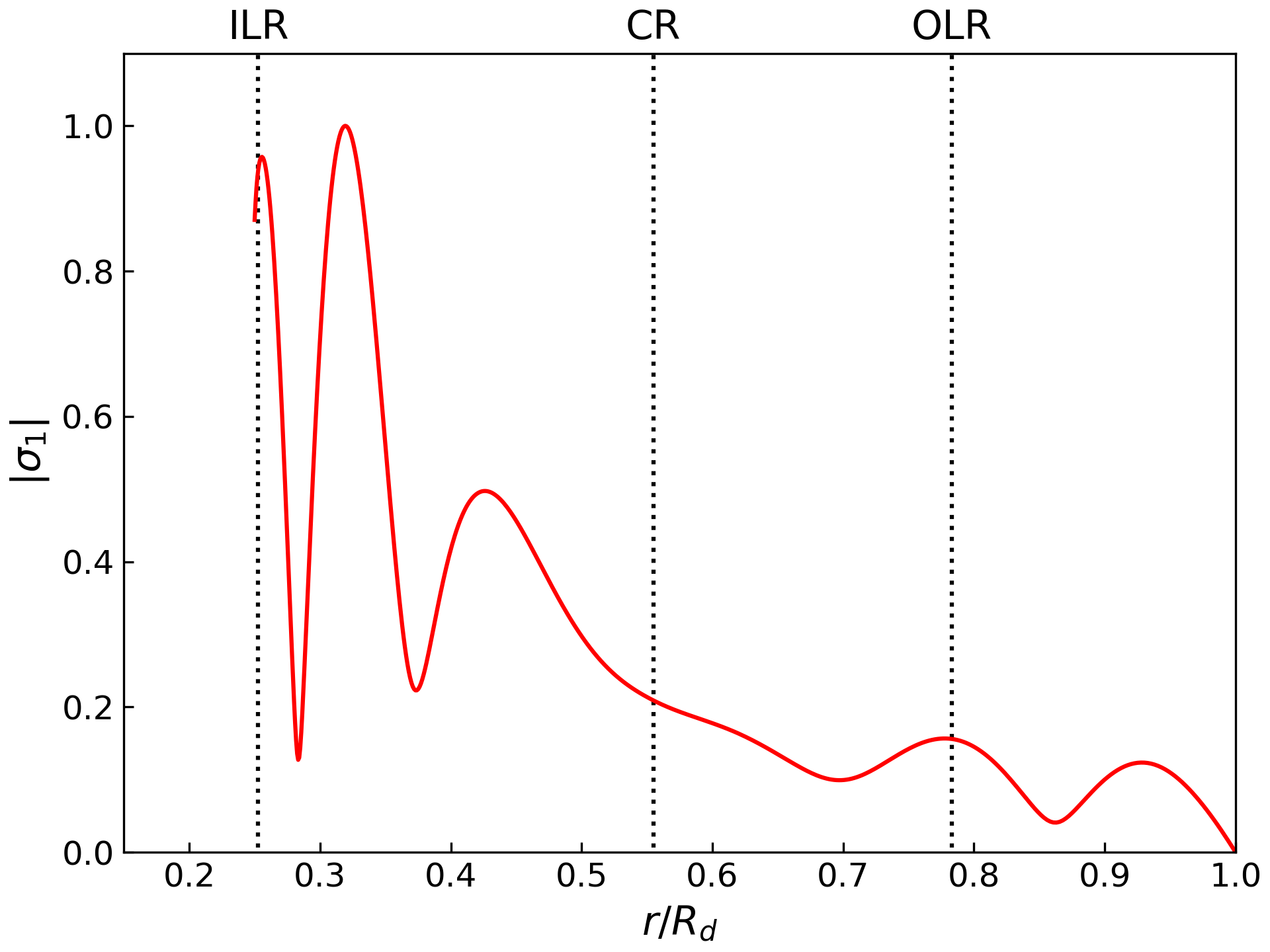}
\caption{Illustration of our calculated spiral density wave mode in the fiducial power law disk. (Left) Two-dimensional distribution of spiral perturbation density on a logarithmic scale, with white curves denoting fitted peak arm position, and inner, middle, and outer dotted circles marking the inner Lindblad (ILR), corotation (CR), and outer Lindblad (OLR) radius, respectively. The spiral pattern between the two Lindblad radii is physically meaningful. (Right) Radial profile of normalized perturbation density $|\sigma_1(r)|$, reflecting the amplitude along the spiral arms in the left panel. \label{Fig:power law fiducial}}
\end{figure*}

\begin{figure*}[h!]
\plottwo{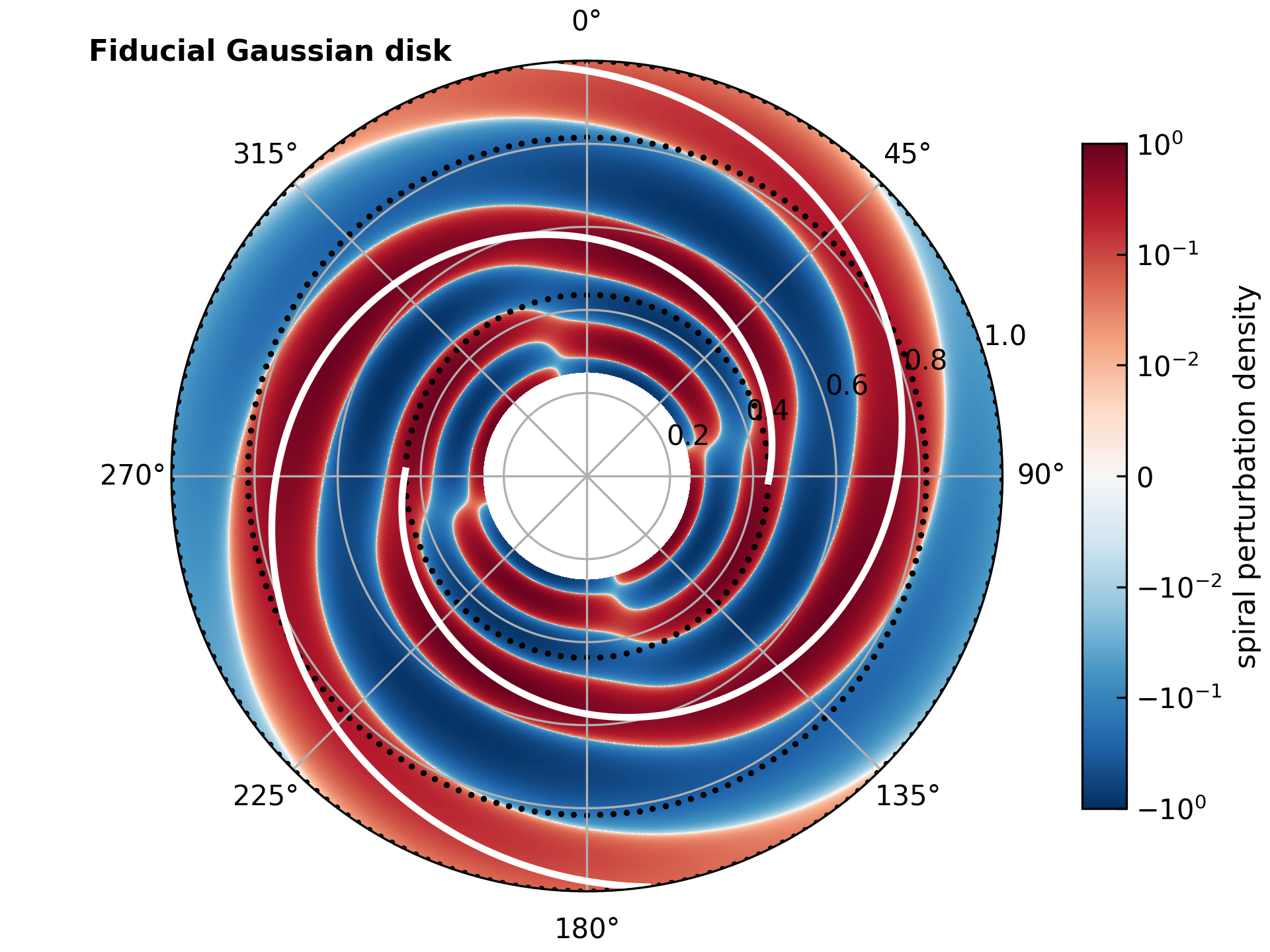}{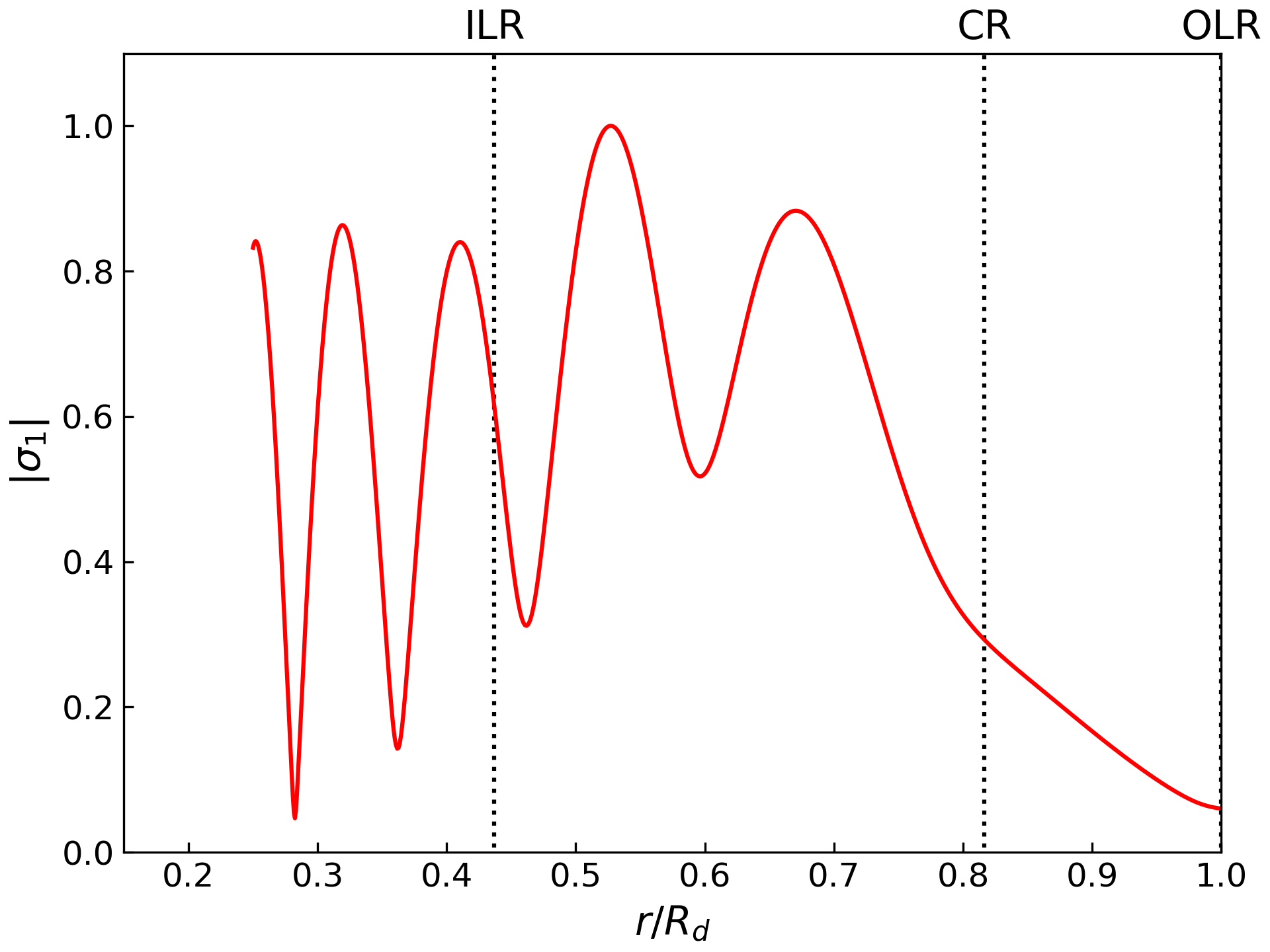}
\caption{Illustration of our calculated spiral density wave mode in the fiducial Gaussian disk, presented in the same way as in Figure~\ref{Fig:power law fiducial}. \label{Fig:Gaussian fiducial}}
\end{figure*}
%---------------------------------------------------------------------

We use the power law disk and the Gaussian disk with $M_d/M_*=0.3$ and $\overline Q=1.40$ as the two fiducial disks of our calculations, which have disk mass and Toomre $Q$ values in the middle of our survey range (Section~\ref{subsec:disk properties}).  The resulting two-armed spiral mode of the fiducial power law disk is illustrated in Figure~\ref{Fig:power law fiducial}. The left panel presents the two-dimensional distribution of spiral density wave  (Re[$\sigma_1(r)e^{-im\phi}$]) on a logarithmic scale, and the right panel shows the amplitude of the perturbation density as a function of radius ($|\sigma_1(r)|$). This mode has pattern speed $\Omega_p=2.61$ and growth rate $\gamma=0.243$ (in units of $\Omega_{\rm K}$, the Keplerian velocity at the disk outer boundary). Figure~\ref{Fig:Gaussian fiducial} presents the results for the fiducial Gaussian disk, with $\Omega_p=1.57$ and $\gamma=0.242$.  The corotation radius [CR; $\Omega(r={\rm CR})=\Omega_p$] marks the radius in which the rotational speed is equal to the pattern speed of spiral arms, and the CR lies between the inner Lindblad radius and the outer Lindblad radius, as marked in Figures~\ref{Fig:power law fiducial} and \ref{Fig:Gaussian fiducial}. In both of the two fiducial disks, the two-dimensional distribution of spiral density wave clearly shows two symmetric, trailing spiral arms, which extend over $270\degr$ in the azimuthal direction.  The quantity $|\sigma_1(r)|$ is the relative amplitude of the perturbation density, normalized by the maximum value, along the ridges of the spiral arms. The amplitude $|\sigma_1(r)|$ fluctuates with radius, as can be seen in both the radial profile and two-dimensional density distribution.  

The principal range for the propagation of spiral density waves lies between the inner Lindblad radius (ILR; $\Omega_p=\Omega-\kappa/m$) and the outer Lindblad radius (OLR; $\Omega_p=\Omega+\kappa/m$) \citep{Bertin1996}. The density waves are absorbed at the Lindblad resonances \citep{LyndenBell1972, Goldreich1978, Goldreich1979}. Therefore, for our study of global spiral density wave modes, we restrict the permitted region for the density waves to the region between the ILR and OLR. As seen in Figures~\ref{Fig:power law fiducial} and \ref{Fig:Gaussian fiducial}, our calculated density wave modes generally have continuous spiral arms with smooth morphology between the two resonances. In contrast, structures outside the ILR and OLR have abnormal appearances.  In the fiducial power law disk (Figure~\ref{Fig:power law fiducial}), a lump piles up between the OLR and the outer boundary, likely due to the imposed outer boundary condition. In the fiducial Gaussian disk (Figure~\ref{Fig:Gaussian fiducial}), structures inside the ILR resemble a set of alternating bananas \citep[also seen in][]{Adams1989}. This structure may arise from the interference of trailing and leading waves with comparable amplitudes. In the principal region where the growth rate of the spiral mode is comparable to the rotation period, the trailing wave dominates over the leading wave and forms continuous spiral arms.

\subsection{Measurement of Spiral Pitch Angle}  \label{subsec:Measurement of Spiral Pitch Angle}

The pitch angle describes the degree of spiral arm winding. Smaller pitch angles correspond to more tightly wound spiral arms, and vice versa.  To measure the pitch angles of our spiral density wave modes, we trace the points of maximum perturbation density (the ridge) at each radius.  Fitting the points with a logarithmic function, $\phi=\beta \ln r+\phi_0$, the pitch angle $\alpha=\arctan(1/\beta)$. We only consider the spiral density waves in the permitted region from the ILR to the OLR, to mitigate possible adverse effects of the imposed boundary conditions.  The best-fit function is illustrated as white curves in Figures~\ref{Fig:power law fiducial} and \ref{Fig:Gaussian fiducial}. Our fiducial power law disk has pitch angle $\alpha=13\fdg 07$, and the fiducial Gaussian disk has $\alpha=9\fdg 61$. Although we restrict to this particular radial range, fitting over the entire disk does not adversely affect our results significantly, yielding $\alpha=12\fdg 13$ for the fiducial power law disk and $\alpha=9\fdg 18$ for the fiducial Gaussian disk. 

%---------------------------------------------------------------------
\begin{figure*}
\plotone{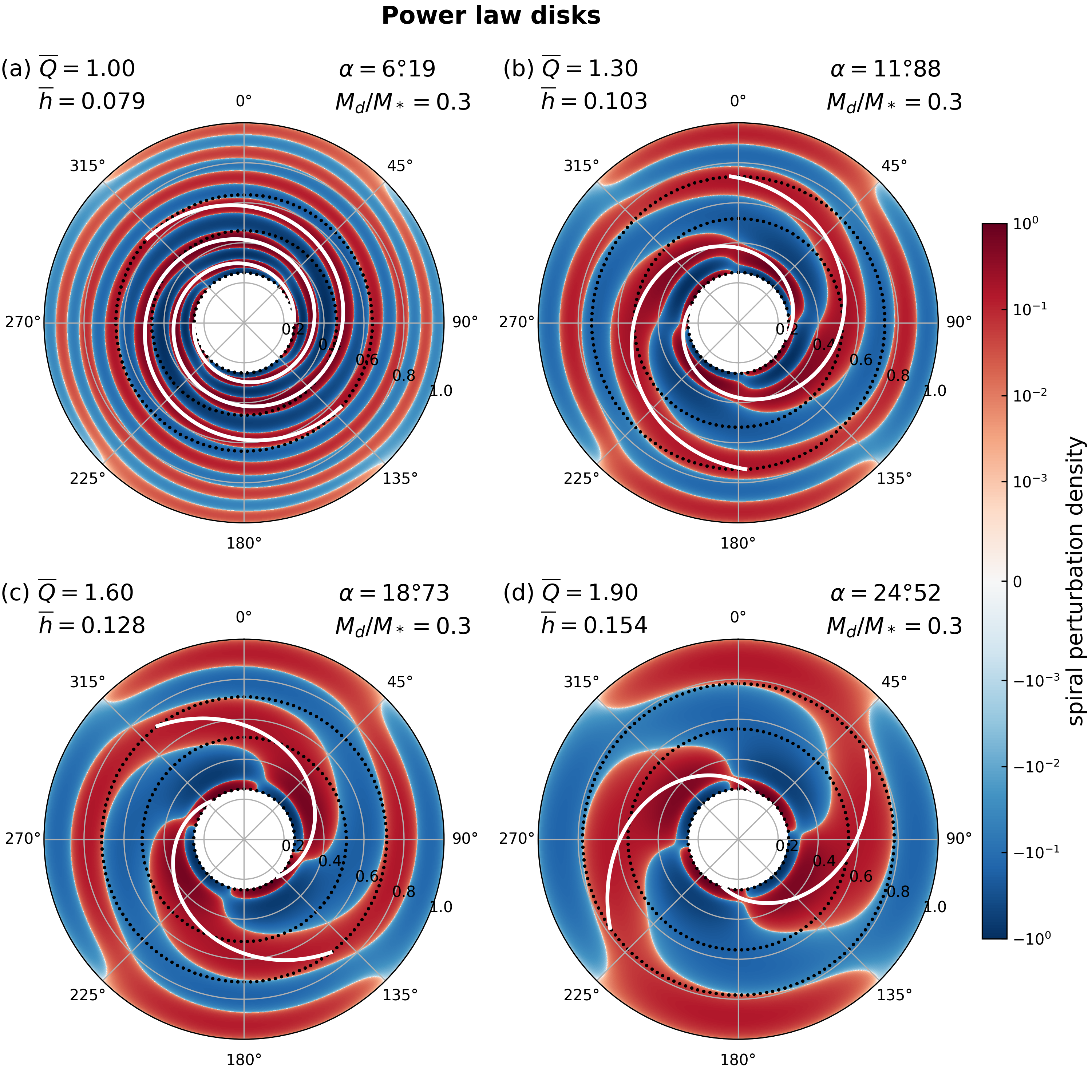}	
\caption{Morphology  of spiral density wave modes in power law disks with fixed $M_d/M_*=0.3$ and 
(a) $\overline Q=1.00$, $\alpha=6\fdg19$; (b) $\overline Q=1.30$, $\alpha=11\fdg88$; (c) $\overline Q=1.60$, $\alpha=18\fdg73$; and (d) $\overline Q=1.90$, $\alpha=24\fdg52$.  The value of $\overline h$ changes accordingly, as $\overline h$ connects with $\overline Q$. \label{Fig:powerdisplay}  }
\end{figure*}

\begin{figure*}
\plotone{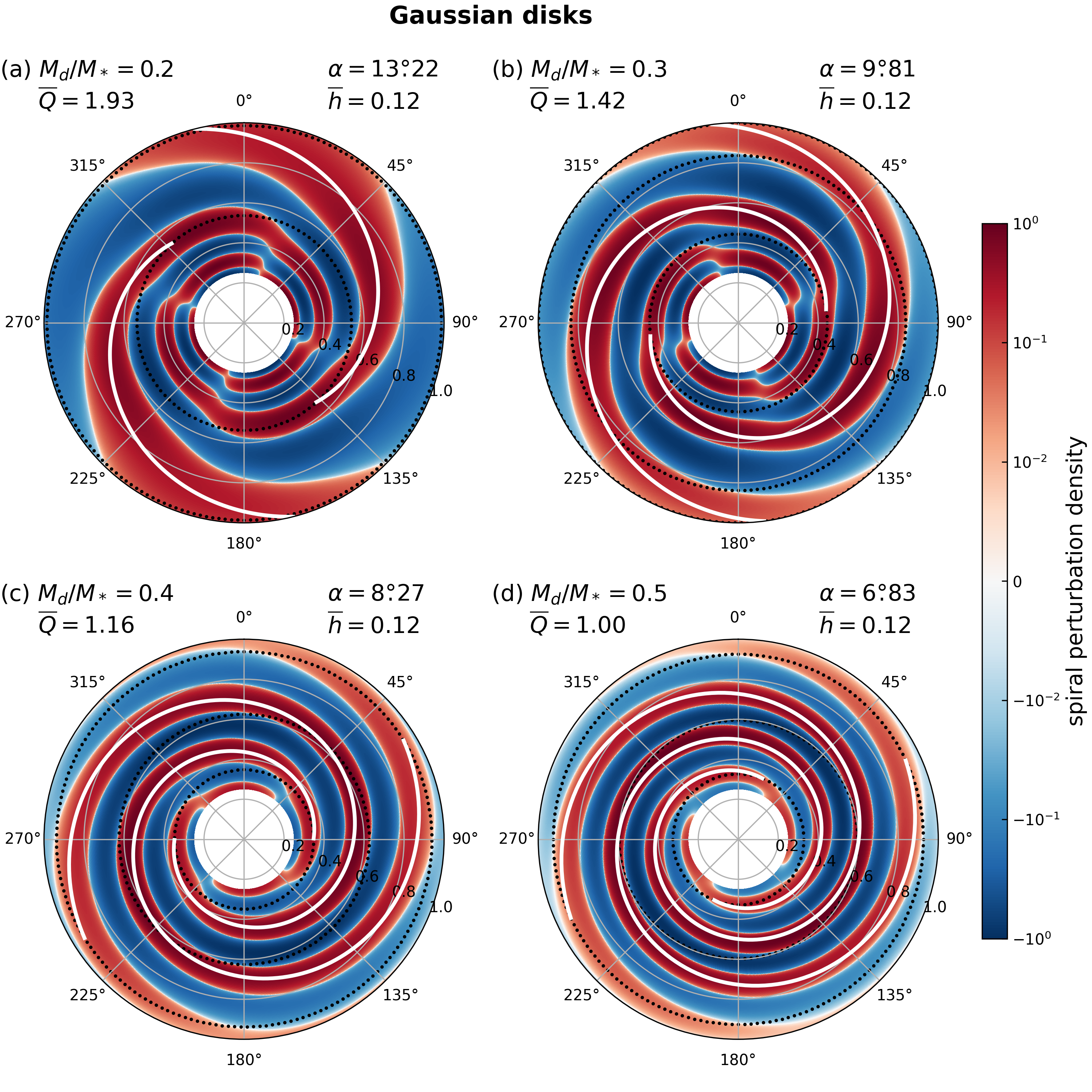}	
\caption{Morphology  of spiral density wave modes in Gaussian disks with fixed $\overline h=0.12$ and (a) $M_d/M_*=0.2$, $\alpha=13\fdg22$; (b) $M_d/M_*=0.3$, $\alpha=9\fdg81$; (c) $M_d/M_*=0.4$, $\alpha=8\fdg27$; and (d) $M_d/M_*=0.5$, $\alpha=6\fdg83$. \label{Fig:Gaussiandisplay}  }
\end{figure*}
%---------------------------------------------------------------------

\subsection{Interdependence Between Spiral Pitch Angle and Disk Properties}  \label{subsec: spiral pitch angle surveys}

In order to understand the behavior of spiral pitch angles, we perform a survey study of two-armed spiral density wave modes on various disk conditions. In our survey, we consider the effect of three parameters: disk-to-star mass ratio ($M_d/M_*$), average Toomre  $Q$ ($\overline Q$), and average aspect ratio ($\overline h$). The surface density profile, temperature profile, and boundary conditions remain the same as for the fiducial disks. The three parameters are connected with each other (see Section~\ref{subsec:disk properties}). We thus have one parameter fixed and one vary, and then observe the variations of spiral pitch angle. We calculate the density wave modes in both power law disks and Gaussian disks.

Figure~\ref{Fig:powerdisplay} displays four wave modes in power law disks with $\overline Q$ increasing from 1 to 1.9 for a given $M_d/M_* = 0.3$. Clear symmetric spiral patterns are present, which become progressively more loosely wound as $\overline Q$ rises. More loosely wound spiral arms cover smaller azimuthal extension, while tightly wound spirals can wind up more than one round. Figure~\ref{Fig:powerdisplay} illustrates examples in Gaussian disks with $M_d/M_*$ increasing from 0.2 to 0.5 for a given $\overline h = 0.12$. The resulting spiral patterns become more tightly wound as $M_d/M_*$ increases. Compared with power law disks, the spirals of Gaussian disks have a larger OLR, which is close to the outer boundary of the disk.

\begin{figure*}[ht]
\epsscale{1}
\plotone{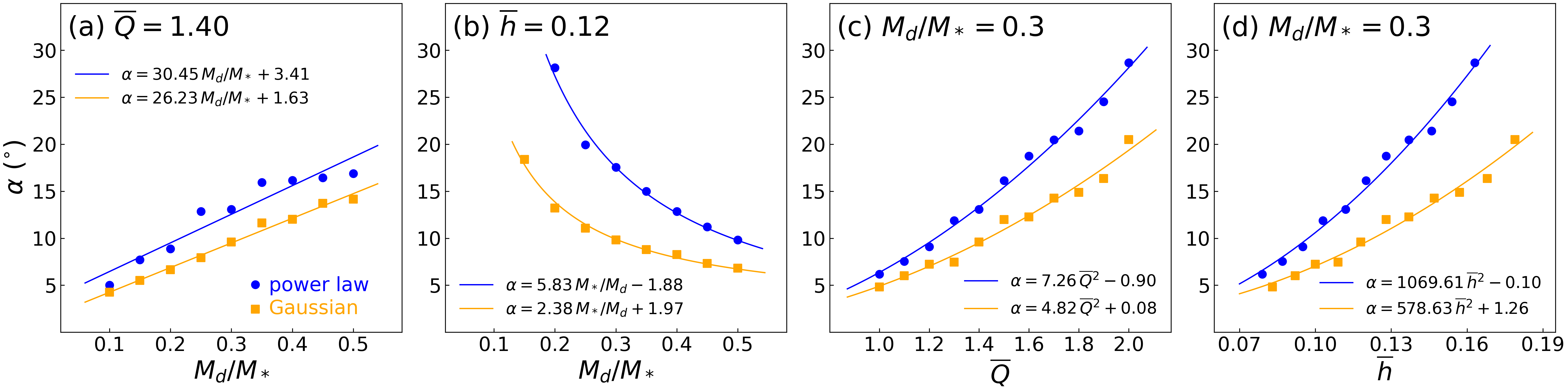}	
\caption{The dependence of the pitch angle ($\alpha$) of spiral density waves in power law disks (blue dots) and Gaussian disks (orange squares) with (a) disk-to-star mass ratio ($M_d/M_*$), with average Toomre $Q$ fixed to $\overline Q = 1.40$, (b) $M_d/M_*$, with average aspect ratio fixed to $\overline h=0.12$, (c) $\overline Q$, with $M_d/M_* = 0.3$, and (d) $\overline h$, with $M_d/M_* = 0.3$. The fitted functions of $\alpha$ and the various parameters are shown as blue (power law) and orange (Gaussian) curves in each panel.\label{Fig:pitch angle survey}}
\end{figure*}

Figure~\ref{Fig:pitch angle survey} presents the interdependence between spiral pitch angles of our density wave modes, measured as in Section~\ref{subsec:Measurement of Spiral Pitch Angle}, and the three main parameters of protoplanetary disks. The blue points and orange squares mark, respectively, the results of power law and Gaussian disks. The pitch angles in our survey, which range from $\alpha \approx 5\degr$ to $30\degr$, exhibit the following trends:

\begin{itemize}
  \item For a fixed  $\overline Q$, a heavier disk (hence larger aspect ratio) has spirals with larger pitch angle (Figure~\ref{Fig:pitch angle survey}a).

  \item For a fixed $\overline h$, a heavier disk is more unstable (smaller $\overline Q$) and has spirals with smaller pitch angle (Figure~\ref{Fig:pitch angle survey}b).

  \item For a fixed $M_d/M_*$, disks with higher $\overline Q$ or higher $\overline h$ show spirals with larger pitch angles (Figures~\ref{Fig:pitch angle survey}c and \ref{Fig:pitch angle survey}d).

  \item Under the same conditions, spirals in a Gaussian disk have smaller pitch angles than those in a power law disk.
\end{itemize}

%===================================================
%===================================================

\section{Discussion}  \label{sec:discussion}

\subsection{Analytical Approach and Implications for the Relationship between Pitch Angle and Disk Properties}   \label{subsec: spiral pitch angle and disk properties}

Our numerical calculations give the relationships between spiral pitch angle and disk properties. To understand the physics behind these trends, we follow \cite{LinShu1964} and employ a widely used dispersion relation of density wave theory to connect the radial wave number with disk properties: 

\begin{equation} \label{eq:dispersion relation}  
(\omega-m\Omega)^2=k^2c_s^2-2\pi G\sigma_0|k|+\kappa^2.
\end{equation}

\noindent
The dispersion relation, derived through the Wentzel-Kramers-Brillouin (WKB) approximation, requires that the phase of the perturbation changes rapidly with radius ($|kr|\gg1$), such that long-range gravitational coupling is negligible and the wave behavior is determined by local dynamics.  Given these assumptions, the dispersion relation works most effectively for tightly wound spiral arms, but still applies to more open and global spirals \citep{Binney2008, Shu2016}. 

The spiral pitch angle $\alpha \approx \partial r/(r\partial\phi) \approx m/(|k|r)$ \citep{Binney2008}. The right-hand side quantity of Equation~(\ref{eq:dispersion relation}) has a minimum value at $|k|=\pi G \sigma_0/c_s^2$, for which the corresponding $\omega$ has the largest growth rate \citep{Cossins2009}. For this most unstable mode, the pitch angle

\begin{equation} \label{eq:pitch}
\alpha \approx \frac{mc_s^2}{\pi G \sigma_0 r}.
\end{equation} 
\

\noindent
Averaging along radius, the pitch angle estimated by Equation~(\ref{eq:pitch}) gives $\alpha=15\fdg44$ for the fiducial power law and $13\fdg80$ for the fiducial Gaussian disk, while our numerical calculation gives $\alpha=13\fdg07$ and $9\fdg61$, respectively. In our disk models, the pitch angle in Equation~(\ref{eq:pitch}) is nearly constant along the radius; for instance, for the power law disk, $\alpha\propto r^{-0.1}$. Because of the weak dependence on radius and the similar disk size in our models, we drop the variable $r$ from the discussions below, and Equation~(\ref{eq:pitch}) becomes

\begin{equation} \label{eq:pitch tendency}
\alpha \propto \frac{c_s^2}{\sigma_0}.
\end{equation} 

\noindent
A denser disk with lower sound speed has more tightly wound spiral arms. If we replace $c_s$ and $\sigma_0$ by $M_d/M_*$, $Q$, or $h$, and consider Keplerian $\Omega$ and $\kappa$, Equation~(\ref{eq:pitch tendency}) yields the following trends:

\begin{enumerate}
\item When $Q$ ($\propto \sqrt{M_*}c_s/M_d$) is fixed, increasing $M_d/M_*$ elevates $c_s$. Because of the quadratic dependence of $c_s$ in Equation~(\ref{eq:pitch tendency}), pitch angle accordingly increases. This results in 
the trend $\alpha\propto {\overline Q}^2 M_d/M_*\propto M_d/M_*$ (Figure~\ref{Fig:pitch angle survey}a).

\item When $h$ ($\propto c_s/\sqrt{M_*}$) is fixed, $c_s$ remains unchanged. Pitch angle decreases as $M_d/M_*$ increases. This results in the trend $\alpha\propto {\overline h}^2 M_*/M_d \propto (M_d/M_*)^{-1}$ (Figure~\ref{Fig:pitch angle survey}b).

\item When $M_d/M_*$ is fixed, increasing of $Q$ or $h$ leads to higher $c_s$, and hence larger pitch angle. This results in the trend $\alpha\propto {\overline Q}^2M_d/M_*\propto {\overline Q}^2$ (Figure~\ref{Fig:pitch angle survey}c) and $\alpha\propto {\overline h}^2M_*/M_d\propto {\overline h}^2$ (Figure~\ref{Fig:pitch angle survey}d).
\end{enumerate}

We further fit the resulting dependence of pitch angle on $M_d/M_*$, $M_*/M_d$, ${\overline Q}^2$, and ${\overline h}^2$ (Figure~\ref{Fig:pitch angle survey}). These new relations, especially that between $\alpha$ and $M_*/M_d$, can successfully fit our numerical results, suggesting that the trends involving the pitch angle of our spiral density wave modes can be explained uniformly by Equation~(\ref{eq:pitch tendency}), in spite that Equation~(\ref{eq:pitch tendency}) involves WKB approximation. Therefore, the reason why Gaussian disks have smaller pitch angles than power law disks is that, between the two Lindblad resonances, for the same total mass there is a broader radial extent in which the surface density of a Gaussian disk is higher than that of a power law disk (Figure~\ref{Fig:density contrast}).  This suggests that more evolved disks should have more tightly wound spirals, as the surface density profiles of evolved disks tend to be close to Gaussian \citep{Laughlin1997, Boehler2017, Pinilla2018}. 

Consistent with the trend of our density wave modes (Figure~\ref{Fig:pitch angle survey}a), the GI-excited spiral arms generated in hydrodynamical simulations of marginally stable ($Q\approx 1$) disks tend to be more open (larger pitch angle) in more massive disks \citep{Cossins2009}.  Evidence supporting the dependence on disk mass and sound speed (Equation~\ref{eq:pitch tendency}) has been found in observational studies. \cite{Yu2019}, studying the spiral pitch angles of 13 protoplanetary disks with a narrow range of disk aspect ratio ($0.06-0.12$), found that more massive disks, that are hence more unstable, have smaller pitch angles, consistent with the trend in Figure~\ref{Fig:pitch angle survey}b. However, the low-mass end of their correlation is less likely to be explained by GI-excited wave modes, as the corresponding disks are too stable.  Near-infrared scattered-light imaging observes the structure on the disk surface, while  millimeter continuum imaging detects the cold dust in the disk midplane. The spiral arms of MWC~758 imaged in near-infrared scattered light are more open than their counterparts observed in the millimeter continuum, which may be due in part to the higher sound speed on the disk surface, but the main cause is the combined geometric effect of a cone-shaped surface and a  flat midplane \citep{Dong2018a}. There is a degree of similarity between the GI-excited spiral density waves studied in this work and planet-excited density wakes.  Similar to our results, planet-excited arms are more open in disks with higher sound speed \citep{Rafikov2002}, in spite of the additional dependence on radius and planet mass \citep{Rafikov2002, Zhu2015}. The pitch angle of planet-excited spirals has similar but weaker dependence on disk mass, if $Q$ is close to unity \citep{ZhangZhu2020}. Inefficient cooling, by raising the sound speed \citep{ZhangZhu2020}, is also expected to impact the pitch angle of our density wave modes in the same manner.  We note that, unlike planet-excited spirals, disks with high $Q$ ($\ga 2$) are unable to generate GI-excited spiral density waves because of the inadequate growth rate.

\subsection{Influence of Boundary Conditions}  \label{subsec: boundary results}

\begin{deluxetable}{c|ccc|ccc}
\tablecaption{Influence of Boundary Conditions \label{table:boundary}}
\tablehead{ 
\colhead{Boundary} & 
\colhead{$\Omega_p^{\rm p}$} & 
\colhead{$\gamma^{\rm p}$} & 
\colhead{$\alpha^{\rm p}$} & 
\colhead{$\Omega_p^{\rm G}$} & 
\colhead{$\gamma^{\rm G}$} & 
\colhead{$\alpha^{\rm G}$} \\
\colhead{Conditions} & 
\colhead{($\Omega_{\rm K}$)} & 
\colhead{($\Omega_{\rm K}$)} & 
\colhead{($\degr$)} & 
\colhead{($\Omega_{\rm K}$)} & 
\colhead{($\Omega_{\rm K}$)} & 
\colhead{($\degr$)}
}
\startdata
RR & 2.56 & 0.136 & 12.92 & 1.11 & 0.468 & 8.85\\
RT & 2.61 & 0.243 & 13.07 & 1.35 & 0.374 & 9.34\\
RC & 2.61 & 0.240 & 13.09 & 1.57 & 0.242 & 9.61\\
TR & 2.32 & 0.193 & 14.30 & 1.11 & 0.483 & 8.76\\
TT & 2.30 & 0.241 & 14.08 & 1.39 & 0.386 & 9.40\\
TC & 2.30 & 0.242 & 14.10 & 1.18 & 0.328 & 8.88\\
\enddata
\tablecomments{Parameters of spiral density wave modes of our fiducial power law (noted with p) and Gaussian (noted with G) disk calculated by adopting different boundary conditions. The first and second letters of the abbreviation in the first column denote the inner and outer boundary conditions: R = reflecting, T = transmitting, and C = confining pressure. The Keplerian rotation speed at the disk outer boundary is $\Omega_{\rm K}$.}
\end{deluxetable}

We investigate the influence of boundary conditions on the density wave modes by altering the boundary conditions adopted in our fiducial disks (see Section~\ref{subsec:BC}).  The parameters of the resulting wave modes are listed in Table~\ref{table:boundary}.  All the wave modes have considerable growth rate and thus exist. The typical fractional change of pitch angle or pattern speed induced by the adoption of different boundary conditions is about 10\%. We conclude that our wave modes are not an artificial result of certain boundary conditions and that the usage of boundary conditions will not significantly influence the derived pitch angle or pattern speed. 

Boundary conditions affect the growth rate \citep[e.g.,][]{Noh1991}.  As Table \ref{table:boundary} illustrates, the growth rates of wave modes with reflecting (R), transmitting (T), and confining pressure (C) inner/outer boundaries generally show the tendency for $\gamma_{\rm R}<\gamma_{\rm T}\approx \gamma_{\rm C}$ for the fiducial power law disk.  By comparison, with the exception of $\gamma_{\rm R}<\gamma_{\rm T}$ for inner boundary, $\gamma_{\rm R}>\gamma_{\rm T}>\gamma_{\rm C}$ for the fiducial Gaussian disk. One possible reason is that the reflected wave at the boundary constructively/destructively interferes, resulting in larger/smaller growth rate than those with a transmitting boundary \citep{Shu2016}.
Another possibility is that, owing to the WKB approximation, our transmitting boundary is not exactly accurate and some reflection may affect the growth rate \citep{Noh1991}. The real boundaries of protoplanetary disks can be combinations of these three idealized boundaries, or perhaps even more complicated.

\subsection{Radial Variation of Spiral Amplitude and Pitch Angle}

In our spiral density wave modes, the amplitude of the arms fluctuates with radius, as can be seen in the  dips and bumps in Figures~\ref{Fig:power law fiducial} and \ref{Fig:Gaussian fiducial}.  GI-excited spirals in numerical simulations exhibit a saturation amplitude with less fluctuation \citep{Cossins2009}. In contrast, the amplitude of planet-launched spirals piles up to a maximum that depends on the planet mass, followed by a decline toward the planet \citep{Zhu2015}.  Recent ALMA observations of spiral arms in protoplanetary disks reveal profiles of fluctuating amplitudes \citep{Huang2018} that resemble those seen in our spiral mode analysis.  In particular, the spiral amplitude profiles of Elias~27, WaOph~6, and IM~Lup have, respectively, one, two, and three main bumps.  Confirming whether they correspond to the density wave modes studied in this work requires detailed modeling, and will be addressed in future work. 

\begin{figure*}[ht]
\plotone{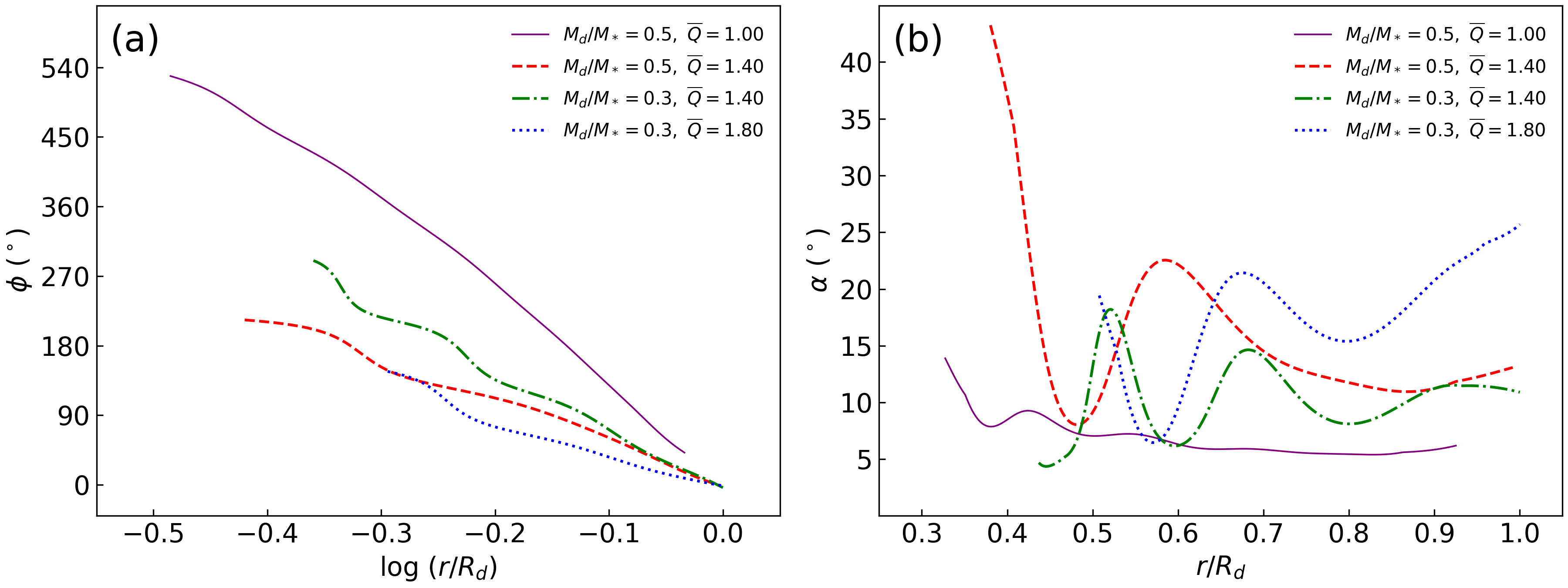}
\caption{Spiral arms azimuthal angle ($\phi$) and pitch angle ($\alpha$) plotted as functions of radius. Results for Gaussian disks with ($M_d/M_*$, $\overline Q$) = $(0.5, 1.00)$, $(0.5, 1.40)$, $(0.3, 1.40)$, and $(0.3, 1.80)$ are marked in purple solid, red dashed, green dashed-dotted, and blue dotted curves, respectively.\label{Fig: pitch angle and radius}}
\end{figure*}

The spiral pitch angle of our calculated density wave modes also varies with radius.  We trace the spiral arms by the points of maximum perturbation density at each radius, and we use the slope of their azimuthal angle $\phi(r)$ to indicate the pitch angle, as $\alpha(r)\approx \partial r/(r\partial \phi)$. We estimate $\alpha(r)$ by fitting a logarithmic function (Section~\ref{subsec:Measurement of Spiral Pitch Angle}) to the ($\phi$, $r$) diagram at each radius,  using a window of 30 points centered on that radius. For the same disk properties, Gaussian disks present a similar radial variation of pitch angle as power law disks (not shown), but with larger extension of continuous spiral arms. We thus focus on the results of $\phi(r)$ and $\alpha(r)$ based on Gaussian disks (Figure~\ref{Fig: pitch angle and radius}); the purple solid, red dashed, green dashed-dotted, and blue dotted curves mark the result for the disk with ($M_d/M_*$, $\overline Q$)$\,=\,$(0.5, 1.00), (0.5, 1.40), (0.3, 1.40), and (0.3, 1.80), respectively.  The radial variation of pitch angle of our density wave modes is related to the disk mass and sound speed. Massive cool disks have pitch angle falling with radius, while low-mass hot disks have pitch angle rising with radius, although oscillations exist in these radial profiles.  Interestingly, the pitch angle of GI-excited spiral arms generated in numerical simulations exhibit little radial variation \citep[e.g.,][]{Cossins2009, Forgan2018}. 

By contrast, the pitch angle of planet-excited spiral wakes increases with radius toward the position of the perturber, beyond which the trend reverses \citep{Rafikov2002, Zhu2015}. We suggest that the radial variation of spiral amplitude and pitch angle potentially provides an effective diagnostic tool to distinguish between planet-excited wakes and GI-excited global density waves.

%===================================================
%===================================================

\subsection{Excitation and Propagation of the Density Wave Mode}  \label{subsec:existence}

Disk mass is a crucial parameter for GI to occur. Spiral arms observed in protoplanetary disks with low mass ($\sim0.01M_*$) are less likely to be excited by GI, and, instead, are probably driven by an embedded planet \citep{Dong2015b, Dong2016b, Fung2015, Zhu2015}. GI-excited spiral arms prefer early-type (Class 0, Class I, and the earliest Class II) protoplanetary disks, which are more massive and fed by infall from their surrounding envelope \citep[e.g.,][]{Dong2016a, Perez2016, Meru2017, Tomida2017}. Nevertheless, it is worth noting that the disk mass, based on the dust mass derived from observations of the millimeter and submillimeter continuum emission, has large uncertainty.  The disk mass can be 
underestimated if the dust is optically thick, grain growth has occurred, or the gas-to-dust mass ratio is larger than assumed \citep{Perez2012,Kratter2016}.  On the other hand, the high stellar accretion rate of Herbig stars indicates that disk masses could be much larger than those inferred from dust emission \citep{Dong2018b}. More accurate measurement of disk masses in the future can help determine whether GI plays an important role in protoplanetary disks.  

A density wave gets absorbed in the Lindblad resonances and thus mainly exists between the ILR and OLR \citep{Adams1989, Bertin1996}.  The density wave will not remain in a fixed position, but propagates radially as a wave packet with a group velocity given by $v_g=\partial \omega(k, r)/\partial k$ \citep{Toomre1969}. We make use of the dispersion relation (Equation~\ref{eq:dispersion relation}) to estimate 

\begin{equation}
  v_g(r) = \pm \frac{|k|c_s^2 - \pi G \sigma_0}{\omega-m\Omega}.
\end{equation}

\begin{figure}
\plotone{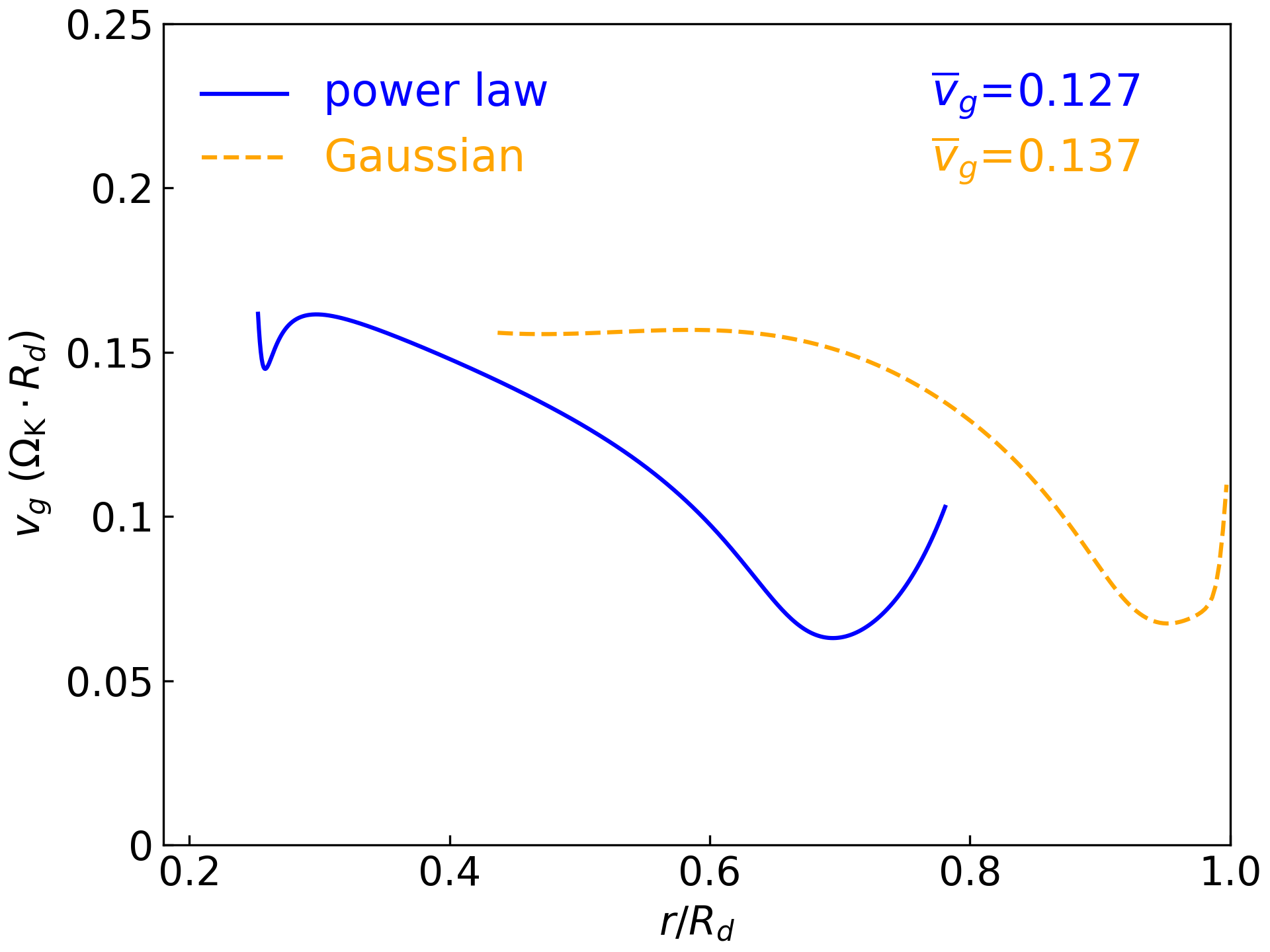}
\caption{Group velocity of spiral density wave modes between inner and outer Lindblad radius. Blue solid and orange dashed curves mark the results in the fiducial power law and Gaussian disk, respectively. The average group velocity is shown at the upper-right corner. 
\label{Fig:group velocity}}
\end{figure}

\noindent
Figure~\ref{Fig:group velocity} plots $v_g(r)$, in units of the Keplerian linear velocity at the disk outer boundary, for the spiral density wave modes of the two fiducial disks.  The average group velocity over the region between the ILR and OLR is $\overline v_g = 0.127$ for the power law disk and $\overline v_g = 0.137$ for the Gaussian disk.  The time scale for the density wave to travel from the ILR to the OLR is thus $t\approx \Delta R/\overline v_g \approx 0.7\,\tau$, where $\tau$ is the rotation period at outer boundary of the disk. When $M_*=0.5\,M_{\odot}$, $t=10^3\,{\rm yr}$. The density wave will have a lifetime of order the orbital time scale, if no feedback or maintenance mechanism is included.  The density wave exists longer if the effect of the ILR is avoided and the wave is amplified by some mechanisms \citep{Bertin1996}.  Short gaseous waves can propagate through the ILR all the way to the center, or through the OLR to infinity \citep{Goldreich1978, Goldreich1979}. For a long wave, the ILR could be avoided if Toomre's $Q$ is sufficiently high in the central region, which acts as a $Q$-barrier to reflect the inward-traveling wave \citep{Lin1970IAU}. If a wave propagates outward and impinges on the corotation zone, it is subject to a turning-point effect, such that the wave evolves into a transmitted wave moving farther out and a stronger reflected wave moving back \citep[overreflection;][]{Mark1974, Mark1976}. The wave is thus amplified.  With these assumptions, ``quasi-stationary'' spiral patterns can exist \citep{Bertin1996}.  Another possible scenario to maintain density waves is that they could act as a perturbation to excite the next generation of density waves, rendering density waves recurrent and similar to the GI-excited spiral arms in numerically simulated massive unstable disks \citep{Tomida2017}. 

There have been many attempts to study GI-excited spiral arms using numerical simulations. It has been shown that disks can settle into a self-regulated state after several rotation periods, after heating by GI-induced dissipation balances cooling by self-radiation \citep{Rice2003, Cossins2009, Rice2009, Forgan2011, Hall2019}.  Simulated spiral structures can also be recurrent, dissipating and then reforming on a time scale of several rotation periods \citep{Nelson1998, LodatoRice2005, Tomida2017}. However, massive, early-type Class 0/I disks can be stabilized shortly after the envelope disperses and accretion from the envelope stops \citep{Tomida2017}.

%====================================================================
%====================================================================

\subsection{Caveats}  \label{subsec:caveats}

This work focuses on $m=2$ modes, motivated by the prevalence of two-armed spirals in recent millimeter observations. In the context of the density wave theory, high-order ($m>2$) modes, having a narrower radial extent between the ILR and OLR, are more likely to be absorbed and therefore less likely to exist. If the conditions to avoid absorption are satisfied, high-order modes can occur and have higher growth rate to replace the $m=2$ mode. In our numerical calculations, the $m = 2,\, 3,\, 4$ modes have a growth rate of 0.243, 0.422, 0.234 in the fiducial power law disk, and 0.242, 0.514, 0.571 in the fiducial Gaussian disk. Numerical simulations of self-gravitating disks have examined the dominant modes with $m\approx M_*/M_d$ \citep{Dong2015a,Hall2019}, somewhat validating our choice of $m=2$ mode for the massive disks probed in this work. While low-mass disks are expected to have many arms, observations of systems such as HD~100453 reveal two arms \citep{Dong2016b}, which might arise from the interaction between the disk and an embedded planet or external star. The nature of spiral arm number has yet to be explored thoroughly with the currently limited observations. Although we focus on $m=2$ modes, their general trends (Figure~6) will hold for high-order modes of multiple arms, should they exist, but with a systematic shift toward larger pitch angles as given by Equation~(\ref{eq:pitch}).

We assume the protoplanetary disk to be razor-thin and analyze the disk dynamics in two dimensions, although the disks in our calculations have aspect ratio around 0.1. Neglecting disk thickness may influence the final results quantitatively, especially in potentially increasing the calculated pitch angle, as the finite thickness of a real disk dilutes the self-gravity \citep{Kratter2016}. Nevertheless, the qualitative physical trends presented in this work will not be adversely affected. In addition, limited to linear analysis, our model does not include shocks, viscosity, accretion, or saturation of wave modes. In the case of planet-excited spirals, the weakly non-linear analysis of \cite{Rafikov2002} is consistent with the linear analysis and simulations by \cite{Zhu2015}, while a new non-linear effect from massive planet is also reported. This suggests that the linear analysis can capture the key behavior of spiral arms, even if the complexities of protoplanetary disks require more detailed hydrodynamical simulations for more complete understanding.

\section{summary}   \label{sec:summary}

Spiral arms have been detected in protoplanetary disks \citep[e.g., ][]{Hashimoto2011, Muto2012, Perez2016, Andrews2018}. An embedded planet or the gravitational instability can excite spiral arms. It has been shown that the pitch angle, a quantity that describes the degree of arm winding, of planet-excited spiral wakes correlates with the disk sound speed, planet mass, and position of the planet \citep{Rafikov2002, Zhu2015, ZhangZhu2020}. However, the behavior of the spiral pitch angle of GI-excited density waves has not been explored. In this work, we calculate and analyze two-armed global spiral density wave modes in thin, non-viscous, self-gravitating protoplanetary disks under the dynamical effect of the gravity from the central star, disk self-gravity, and thermal pressure.  We only consider massive ($M_d/M_*\ge 0.1$), cool (Toomre $\overline{Q} \le 2$) disks, as low-mass hot systems are stable to non-axisymmetric spiral perturbation. Our disk models employ observationally motivated physical conditions, including both power law and Gaussian density profiles, a power law temperature profile, and three sets (reflecting, transmitting, and confining pressure) of boundary conditions. We probe the morphology of the resulting two-armed density wave modes and investigate the influence of the disk-to-star mass ratio ($M_d/M_*$), average Toomre $Q$ ($\overline Q$), and average aspect ratio ($\overline h$) on spiral pitch angle. Our main findings are:

\begin{enumerate}

\item The pitch angle and pattern speed of spiral density wave modes are insensitive to the boundary conditions adopted.
 
\item All else being equal, Gaussian disks possess more tightly wound spiral density waves than power law disks, suggesting that more evolved disks have more tightly wound spirals than less evolved disks.
  
\item At a fixed $\overline Q$, more massive disks, owing to their higher temperatures, have more loosely wound (more open) spirals.
 
\item At a fixed $\overline h$, more massive disks, owing to their higher self-gravity, have more tightly wound spirals.
  
\item At a fixed $M_d/M_*$, disks with higher $\overline Q$ or $\overline h$ have hotter temperatures and thus generate more loosely wound (more open) spirals.
   
\item  The pitch angle scales with sound speed $c_s$ and mass $M_d$ of the disk as $\alpha\propto c_s^2/M_d$.

\item Unlike planet-launched spiral wakes, the amplitude of GI-excited spiral density waves fluctuates with radius, consistent with recent ALMA observations.

\item Unlike planet-launched spiral wakes, the radial variation of pitch angle of GI-excited density waves depends on 
$c_s$ and $M_d$. The spiral pitch angle of massive cool disks decreases with increasing radius, while that of low-mass hot disks increases with radius.
  
\item Spiral density waves propagate at the group velocity on a time scale of a few rotation periods, suggesting that density waves need to be constantly replenished.
 
\end{enumerate}

\noindent
Comparison of our derived behavior of spiral amplitude and pitch angle with observations potentially can test the GI-excited density wave mechanism and constrain the properties of protoplanetary disks.

\acknowledgments{
We acknowledge helpful discussions with Wing-Kit Lee, Zhaohuan Zhu, and Shangjia Zhang. We also thank the referee Ruobing Dong for helpful feedback. This work was supported by the National Science Foundation of China (11721303, 11991052) and the National Key R\&D Program of China (2016YFA0400702). }

\appendix

\section{Numerical Process}  \label{app:numerical}

We follow the numerical method described in \citet{Adams1989} to solve the integro-differential equation (Equation~\ref{eq:int-diff}). We summarize the procedure below. After setting radial grid points on a logarithmic scale, we discretize the integro-differential equation and boundary conditions into matrix form, which gives $\mathcal{W}(\omega)\emph{\textbf{S}}=0$. Equation~(\ref{eq:int-diff}) corresponds to the middle rows of $\mathcal{W}(\omega)$, and the inner and outer boundary condition correspond to the first and last row, respectively.

Equation~(\ref{eq:int-diff}) is transformed into 

\begin{equation}   \label{eq:matrix general}
\begin{aligned}
 (\mathcal{W}_g)_{ik}= &\nu(1-\nu^2)\left \{ \mathcal D^{(2)}_{ij} +\left(Ar+1+\frac{2r}{\sigma_0}\frac{\mathrm{d}\sigma_0}{\mathrm{d}r} \right) \mathcal D^{(1)}_{ij}+\left [ Ar\left(1+\frac{r}{\sigma_0}\frac{\mathrm{d}\sigma_0}{\mathrm{d}r}\right)+Br^2+\frac{r^2}{\sigma_0}\frac{\mathrm{d}^2\sigma_0}{\mathrm{d}r^2}+\frac{2r}{\sigma_0}\frac{\mathrm{d}\sigma_0}{\mathrm{d}r}\right]\delta_{ij} \right \} \mathcal{I}_{jk}
\\&+\frac{c_s^2\nu(1-\nu^2)}{2\pi G \sigma_0 r}\left\{\mathcal D^{(2)}_{ik}+\left[Ar+\frac{2r}{c_s^2}\frac{\mathrm{d}(c_s^2)}{\mathrm{d}r}-1\right]\mathcal D^{(1)}_{ik}+\left[\frac{r^2}{c_s^2}\frac{\mathrm{d}^2(c_s^2)}{\mathrm{d}r^2}+Ar\frac{r}{c_s^2}\frac{\mathrm{d}(c_s^2)}{\mathrm{d}r}+Br^2\right]\delta_{ik}\right\}-\left [ \frac{\kappa^2\nu(1-\nu^2)^2r}{2\pi G\sigma_0} \right]\delta_{ik}, 
\end{aligned}
\end{equation}

\noindent
where $\mathcal{W}_g$ is an $(N-2)\times N$ matrix. $\mathcal D^{(1)}$ and $\mathcal D^{(2)}$ are the first and second order derivation operator, while $\mathcal I$ is the disk self-gravity potential integration in matrix form. The matrix forms of the reflecting (Equation~\ref{eq:matrix reflecting}), transmitting (Equation~\ref{eq:matrix transmitting}), and confining pressure (Equation~\ref{eq:matrix confining pressure}) boundaries read as:

%%%%%%%%%%%%%%% Eq. A2
\begin{equation}  \label{eq:matrix reflecting}
\begin{aligned}
(\mathcal{W}_r)_{ik}=&\left\{(\omega-m\Omega)\mathcal D^{(1)}_{ij}+\left[(\omega-m\Omega)\left(\frac{r}{\sigma_0}\frac{\mathrm{d}\sigma_0}{\mathrm{d}r}+1\right)-2m\Omega\right]\delta_{ij}\right\}\mathcal{I}_{jk}
\\&+\frac{c_s^2}{2\pi G \sigma_0 r}\left\{(\omega-m\Omega)\mathcal D^{(1)}_{ik}+\left[(\omega-m\Omega)\frac{r}{c_s^2}\frac{\mathrm{d}(c_s^2)}{\mathrm{d}r}-2m\Omega\right]\delta_{ik}\right\},
\end{aligned}
\end{equation}
%%%%%%%%%%%%%%% Eq. A3

\begin{subequations}  \label{eq:matrix transmitting}
\begin{equation}  
(\mathcal{W}_{t})_{ik}= \mathcal D^{(1)}_{ik} +r\left(\frac{1}{\sigma_0}\frac{\mathrm{d}\sigma_0}{\mathrm{d}r}+ i|k|\right)\delta_{ik},
\end{equation}

\begin{equation}  
|k|=\frac{\pi G \sigma_0}{c_s^2} + \frac{\pi G \sigma_0}{c_s^2}[1-Q^2(1-m^2)]^{1/2}\quad (\rm{inner~boundary:~} \omega \ll \Omega, \kappa\simeq\Omega),
\end{equation}

\begin{equation}  
|k|=\frac{\pi G \sigma_0}{c_s^2}\left(1-\frac{Qm\Omega}{\kappa}  \right) + \frac{\pi G \sigma_0}{c_s^2} \frac{Q}{\kappa}\omega  \quad (\rm{outer~boundary:~} Q\approx 1),
\end{equation}
\end{subequations}
%%%%%%%%%%%%%%% Eq. A4

\begin{equation}  \label{eq:matrix confining pressure}
\begin{aligned}
(\mathcal{W}_c)_{ik}=&\left\{(\omega-m\Omega)\mathcal D^{(1)}_{ij}+\left[(\omega-m\Omega)\left(\frac{r}{\sigma_0}\frac{\mathrm{d}\sigma_0}{\mathrm{d}r}+1\right)-2m\Omega\right]\delta_{ij}\right\}\mathcal{I}_{jk}
\\&+\frac{c_s^2}{2\pi G \sigma_0 r}\left\{(\omega-m\Omega)\mathcal D^{(1)}_{ik}+\left[ (\omega-m\Omega)\frac{r}{c_s^2}\frac{\mathrm{d}(c_s^2)}{\mathrm{d}r}-2m\Omega\right]\delta_{ik}\right\}-\frac{\kappa^2(1-\nu^2)}{2\pi G \mathrm{d}\sigma_0/\mathrm{d}r}(\omega-m\Omega)\delta_{ik}.
\end{aligned}
\end{equation}
%%%%%%%%%%%%%%%

\noindent
Note that we have made approximations on the wave number $|k|$ of the transmitting boundary, in order for the resulting matrix be of integral order $\omega$. To transform Equation~(\ref{eq:int-diff matrix}) into Equation~(\ref{eq:eigen}), we analytically rewrite $\mathcal{W}(\omega)$ into a fifth-order polynomial of $\omega$, and so Equation~(\ref{eq:int-diff matrix}) becomes

\begin{equation}      \label{eq:polynomial}
[\mathcal{W}^{(5)}\omega^5+\mathcal{W}^{(4)}\omega^4+\mathcal{W}^{(3)}\omega^3+\mathcal{W}^{(2)}\omega^2+\mathcal{W}^{(1)}\omega+\mathcal{W}^{(0)}]\emph{\textbf{S}}=0.
\end{equation}

\noindent
By defining the $5N$-dimensional vector $\emph{\textbf{S}}_\emph{\textbf{d}}(\emph{\textbf{S}},\omega)$, we rewrite Equation~(\ref{eq:polynomial}) into Equation~(\ref{eq:eigen}), in which $\mathcal{T}$ and $\mathcal{Z}$ consist of polynomial coefficients $\mathcal{W}^{(0)}, \mathcal{W}^{(1)},...,\mathcal{W}^{(5)}$. Their expressions can be found in Appendix~B of \citet{Adams1989}.

We use a softened gravity to deal with the singularity that arises in the potential integration (Section~\ref{subsec:numerical method}). The softening parameter $\eta$ should not be too large, which would make the result deviate from real gravity, or too small, which would introduce large numerical errors.  We have tested the influence of $\eta$ on the calculated density wave modes and adopted an appropriate value. For example, for our fiducial power law disk, we use $\eta_1^2=10^{-2}$ for $\Omega$ (a relatively large $\eta$ to assure that the $\Omega$ and $\kappa$ curves are smooth), and we set $\eta_2^2=10^{-5}$ for $\Psi_1$. When we change $\eta_2^2$ to $10^{-7}$, the resulting growth rate changes from 0.243 to 0.223; thus, the softening parameters we chose lie in the convergence region. We have also tested the effect of the number of grid points ($N$) on our results. In the fiducial Gaussian disk, calculation with $N=500$ results in $\omega=1.57-0.242\,i$, while that with $N=1000$ gives almost the same result, $\omega=1.57-0.245\,i$. We thus adopt $N=500$ to improve calculation efficiency.

%===============================================================================
\begin{figure}
\epsscale{0.7}
\figurenum{A1}
\plotone{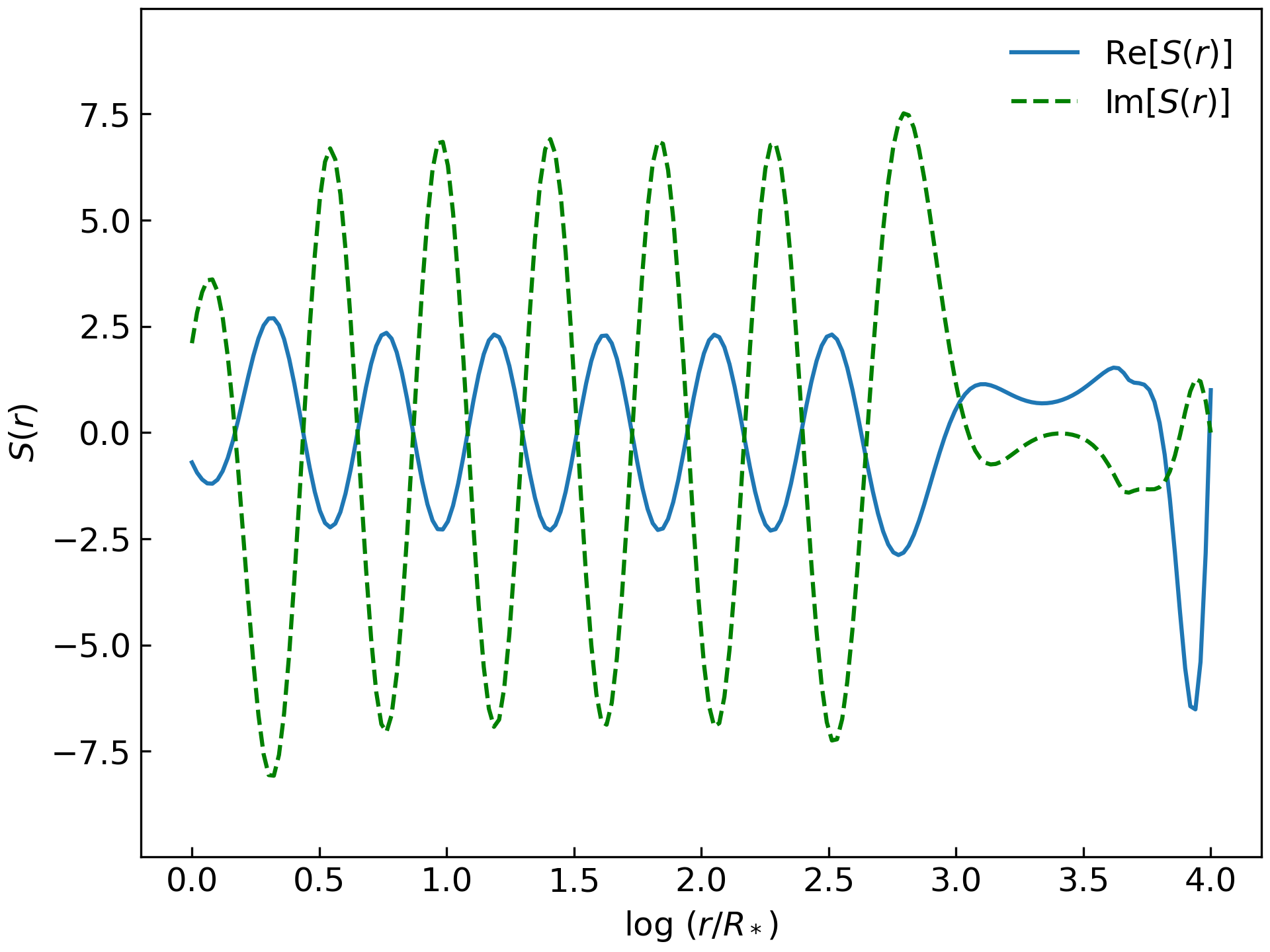}
\caption{Reproduction of Figure~3 of \cite{Adams1989}. This plot is the non-dimensional perturbation density distribution $S(r)=\sigma_1(r)/\sigma_0(r)$ of the lowest mode in the fiducial disk of \cite{Adams1989}. The blue solid line shows the real part of the function, and the green dashed line shows the imaginary part of the function.\label{Fig:reproduce}}
\end{figure}
%===============================================================================

To test the validity of our numerical processing, we reproduce a representative  mode in \citet{Adams1989} (Figure~\ref{Fig:reproduce}).  Our calculated wave frequency ($\omega=4.11-0.217i$) is well consistent with their result ($\omega=4.26-0.232i$), with a difference of only $\sim$5\%. Our perturbation density $S(r)$ is also in excellent agreement with \citet[][see their Figure~3]{Adams1989}. The small difference between the results comes from the different methods we use in calculating $\Omega$ and dealing with the singularity.

\end{document}